\documentclass[10pt,letterpaper]{article}
\usepackage[]{fontenc}
\usepackage[latin9]{inputenc}
\usepackage{geometry}
\usepackage{amsmath}
\usepackage{setspace}
\usepackage{amssymb}
\usepackage[unicode=true, 
 bookmarks=true,bookmarksnumbered=true,bookmarksopen=true,bookmarksopenlevel=3,
 breaklinks=false,pdfborder={0 0 0},backref=false,colorlinks=false]
 {hyperref}
\hypersetup{pdftitle={The Near-Flat-Space and BMN Limits for Strings in AdS4 x CP3 at One Loop},
 pdfauthor={Michael C. Abbott (TIFR) and Per Sundin (UCT)}}

\makeatletter
\pdfoutput=1

\usepackage[T1]{fontenc}
\usepackage{ae}
\usepackage{aecompl}

\usepackage[sf,bf,small]{titlesec}
\usepackage[small,sf,bf,hang]{caption}

\usepackage[multiple]{footmisc}
\long\def\symbolfootnote[#1]#2{\begingroup%
\def\thefootnote{\fnsymbol{footnote}}\footnote[#1]{#2}\endgroup}

\usepackage{titletoc}
\dottedcontents{section}[6mm]{\smallskip }{6mm}{0pc} 
\titlecontents*{subsection}[6mm]{\itshape\small }{\thecontentslabel \ }{}{, \thecontentspage}[ \ \textbullet \ \ ][] 
\def\tableofcontents{\subsection*{\contentsname}\vspace{-2mm}\@starttoc{toc}}

\usepackage[nosort]{cite} 

\usepackage{feynmp}  
\DeclareGraphicsExtensions{.pdf}  % include pdf files by default
\renewcommand{\bar}[1]{\overline{#1}}
\newcommand{\pchain}{p_\mathrm{chain}}

\newcommand{\nn}{\nonumber}

\newcommand{\ns} \normalsize

\newcommand{\PI}{\text{\Large{$\pi$}}}

\AtBeginDocument{
  
}

\makeatother

\begin{document}
%\selectlanguage{english}
\begin{flushright}
TIFR/TH/11-25\vspace{8mm}
\par\end{flushright}

\begin{center}
\textsf{\textbf{\Large The Near-Flat-Space and BMN Limits for \smallskip\smallskip }}\\
\textsf{\textbf{\Large Strings in $AdS_{4}\times CP^{3}$ at One
Loop }}
\par\end{center}{\Large \par}

\begin{singlespace}
\begin{center}
Michael C. Abbott$^{1}$ and Per Sundin$^{2}$ \bigskip \\
{\small $^{1}$}\emph{\small{} Department of Theoretical Physics,}\\
\emph{\small Tata Institute of Fundamental Research, }\\
\emph{\small Homi Bhabha Rd, Mumbai 400-005, India}\\
\emph{\small abbott@theory.tifr.res.in }{\small \bigskip }\emph{\small }\\
{\small $^{2}$}\emph{\small{} Astrophysics, Cosmology \& Gravity
Centre }\\
\emph{\small and Department of Applied Mathematics, }\\
\emph{\small University of Cape Town,}\\
\emph{\small Private Bag, Rondebosch, 7700, South Africa}{\small }\\
\emph{\small nidnus.rep@gmail.com}
\par\end{center}{\small \par}
\end{singlespace}

\begin{center}
3 June 2011
\par\end{center}

\subsection*{\hspace{9mm}Abstract}
\begin{quote}
This paper studies type IIA string theory in $AdS_{4}\times CP^{3}$
in both the BMN limit and the Maldacena--Swanson \cite{Maldacena:2006rv}
or near-flat-space limit. We derive the simpler Lagrangian for the
latter limit by taking a large worldsheet boost of the BMN theory.
We then calculate one-loop corrections to the correlators of the various
fields using both theories. In all cases the near-flat-space results
agree with a limit of the BMN results, providing evidence for the
quantum consistency of this truncation. The corrections can also be
compared to an expansion of the exact dispersion relation, known from
integrability apart from one interpolating function $h(\lambda)$.
Here we see agreement with the results of McLoughlin, Roiban \& Tseytlin
\cite{McLoughlin:2008he}, and we observe that it does not appear
to be possible to fully implement the cutoff suggested by Gromov \&
Mikhaylov \cite{Gromov:2008fy}, although for some terms we can do
so. In both the near-flat-space and BMN calculations there are some
extra terms in the mass shifts which break supersymmetry. These terms
are extremely sensitive to the cutoff used, and can perhaps be seen
as a consequence of using dimensional regularisation. 

\bigskip  \thispagestyle{empty} \nocite{Sundin:2009zu}
\end{quote}

%\input{diagrams.tex}  %the name causes problems on arxiv
% pasted in below instead:

%%
%% This file contains descriptions of feynman diagrams for use with the package feynMP
%% After running the main file, run "mpost diagrams" once to generate these
%%
%%
%% These are for propagators paper with Per. 
%% The version in this folder edited 18 May 2011 to add unlabled ones for introduction. 
%%

\newsavebox{\feynmanrules}
\sbox{\feynmanrules}{
\begin{fmffile}{diagrams} % I can't seem to make this work using any path but the same one as the document

%%%%%%%%%%%%%%%%
%%  SETTINGS

\fmfset{thin}{0.6pt}  % was 0.7 until v24
%\fmfset{wiggly_len}{5mm}
\fmfset{dash_len}{4pt}
\fmfset{dot_size}{1thick}
\fmfset{arrow_len}{6pt} % you can't use em here, mpost doesn't know what it will be.
%\fmfset{curly_len}{2.5mm}
%\setlength{\unitlength}{1em} % default is =1pt, maybe that's sensible. 72pt = 1in

%%%%%%%%%%%%%%%%%%%%
%% TADPOLE GRAPHS  
  
\begin{fmfgraph*}(72,25)
\fmfkeep{tadpole}
\fmfleft{in,p1}
\fmfright{out,p2}
\fmfdot{c}
\fmf{dashes_arrow,label=\small{$\omega_\alpha (p)$}}{in,c}
\fmf{dashes_arrow}{c,out}
\fmf{plain_arrow,right, tension=0.8, label=\small{$\omega_\beta(k)$}}{c,c}
\fmf{phantom, tension=0.2}{p1,p2}
\end{fmfgraph*}

\begin{fmfgraph*}(72,25)
\fmfkeep{tadpole-heavy}
\fmfleft{in,p1}
\fmfright{out,p2}
\fmfdot{c}
\fmf{dashes_arrow,label=\small{$ \omega_\alpha (p)$}}{in,c}
\fmf{dashes_arrow}{c,out}
\fmf{dbl_plain,right, tension=0.8, label=\small{$y(k)\text{ or }z_i(k)$}}{c,c}
\fmf{phantom, tension=0.2}{p1,p2}
\end{fmfgraph*}

\begin{fmfgraph*}(72,25)
\fmfkeep{tadpole-fermi}
\fmfleft{in,p1}
\fmfright{out,p2}
\fmfdot{c}
\fmf{dashes_arrow,label=\small{$\omega_\alpha (p)$}}{in,c}
\fmf{dashes_arrow}{c,out}
\fmf{plain_arrow,right, tension=0.8, label=\small{$\psi^b (k)$}}{c,c}
\fmf{phantom, tension=0.2}{p1,p2}
\end{fmfgraph*}

%\begin{fmfgraph*}(72,25)
%\fmfkeep{tadpole-heavy-less}
%\fmfleft{in,p1}
%\fmfright{out,p2}
%\fmfdot{c}
%\fmf{dashes_arrow,label=\small{$ \omega_\alpha (p)$}}{in,c}
%\fmf{dashes_arrow}{c,out}
%\fmf{dbl_plain,right, tension=0.8, label=\small{ \  }}{c,c}
%\fmf{phantom, tension=0.2}{p1,p2}
%\end{fmfgraph*}

%%% y ext

\begin{fmfgraph*}(72,25)
\fmfkeep{tadpole-yy}
\fmfset{dash_len}{6pt} % this seems to be a local change
\fmfleft{in,p1}
\fmfright{out,p2}
\fmfdot{c}
\fmf{dbl_dashes,label=\small{$ y $}}{in,c}
\fmf{dbl_dashes}{c,out}
\fmf{plain_arrow,right, tension=0.8, label=\small{$\omega_\alpha\text{ or }\psi^a$}}{c,c}
\fmf{phantom, tension=0.2}{p1,p2}
\end{fmfgraph*}

\begin{fmfgraph*}(72,25)
\fmfkeep{tadpole-yy-heavy}
\fmfset{dash_len}{6pt} % this seems to be a local change
\fmfleft{in,p1}
\fmfright{out,p2}
\fmfdot{c}
\fmf{dbl_dashes,label=\small{$ y $}}{in,c}
\fmf{dbl_dashes}{c,out}
\fmf{dbl_plain,right, tension=0.8, label=\small{$y\text{ or }z_i$}}{c,c}
\fmf{phantom, tension=0.2}{p1,p2}
\end{fmfgraph*}

%%% psi ext

\begin{fmfgraph*}(72,25)
\fmfkeep{psi-tad-w}
\fmfleft{in,p1}
\fmfright{out,p2}
\fmfdot{c}
\fmf{dashes_arrow,label=\small{$\psi^a (p)$}}{in,c}
\fmf{dashes_arrow}{c,out}
\fmf{plain_arrow,right, tension=0.8, label=\small{$\omega_\beta(k)$}}{c,c}
\fmf{phantom, tension=0.2}{p1,p2}
\end{fmfgraph*}

\begin{fmfgraph*}(72,25)
\fmfkeep{psi-tad-yz}
\fmfleft{in,p1}
\fmfright{out,p2}
\fmfdot{c}
\fmf{dashes_arrow,label=\small{$\psi^a (p)$}}{in,c}
\fmf{dashes_arrow}{c,out}
\fmf{dbl_plain,right, tension=0.8, label=\small{$y(k)\text{ or }z_i(k)$}}{c,c}
\fmf{phantom, tension=0.2}{p1,p2}
\end{fmfgraph*}

\begin{fmfgraph*}(72,25)
\fmfkeep{psi-tad-psi}
\fmfleft{in,p1}
\fmfright{out,p2}
\fmfdot{c}
\fmf{dashes_arrow,label=\small{$\psi^a (p)$}}{in,c}
\fmf{dashes_arrow}{c,out}
\fmf{plain_arrow,right, tension=0.8, label=\small{$\psi^b (k)$}}{c,c}
\fmf{phantom, tension=0.2}{p1,p2}
\end{fmfgraph*}

\begin{fmfgraph*}(72,25)
\fmfkeep{psi-tad-s}
\fmfleft{in,p1}
\fmfright{out,p2}
\fmfdot{c}
\fmf{dashes_arrow,label=\small{$\psi^a (p)$}}{in,c}
\fmf{dashes_arrow}{c,out}
\fmf{dbl_plain,right, tension=0.8, label=\small{$s^b_\beta (k)$}}{c,c}
\fmf{phantom, tension=0.2}{p1,p2}
\end{fmfgraph*}

%%% psi ext compact
%

%\begin{fmfgraph*}(72,25)
%\fmfkeep{psi-tad-light}
%\fmfleft{in,p1}
%\fmfright{out,p2}
%\fmfdot{c}
%\fmf{dashes_arrow,label=\small{$\psi^a (p)$}}{in,c}
%\fmf{dashes_arrow}{c,out}
%\fmf{plain_arrow,right, tension=0.8, label=\small{$\omega\text{ or }\psi$}}{c,c}
%\fmf{phantom, tension=0.2}{p1,p2}
%\end{fmfgraph*}

%
%\begin{fmfgraph*}(72,25)
%\fmfkeep{psi-tad-heavy}
%\fmfleft{in,p1}
%\fmfright{out,p2}
%\fmfdot{c}
%\fmf{dashes_arrow,label=\small{$\psi^a (p)$}}{in,c}
%\fmf{dashes_arrow}{c,out}
%\fmf{dbl_plain,right, tension=0.8, label=\small{$y\text{, }z\text{ or }s$}}{c,c}
%\fmf{phantom, tension=0.2}{p1,p2}
%\end{fmfgraph*}

%%% lollipop

\begin{fmfgraph*}(72,36)
\fmfkeep{lollipop}

\fmfstraight
\fmfleft{in,i1,i2}
\fmfright{out,o1,o2}
\fmfdot{bot,mid}

\fmf{dashes_arrow}{in,bot} %  removed ,label=\small{$\omega_\alpha (p)$}
\fmf{dashes_arrow}{bot,out}

\fmf{phantom}{i2,top}
\fmf{phantom}{top,o2}
\fmffreeze

\fmf{phantom}{i1,mid}
\fmf{phantom}{mid,o1}

\fmf{dbl_plain, tension=2}{bot,mid}

\fmf{plain_arrow,right, tension=0.8}{mid,top}
\fmf{plain,right, tension=0.8}{top,mid}
\end{fmfgraph*}

%%%%%%%%%%%%%%%%%%%%
%% BUBBLE GRAPHS  

\begin{fmfgraph*}(100,36)
\fmfkeep{bubble}
\fmfleft{in}
\fmfright{out}
\fmfdot{v1}
\fmfdot{v2}
\fmf{dashes_arrow,label=\small{$\omega_\alpha (p)$}}{in,v1}
\fmf{dashes_arrow}{v2,out}
\fmf{plain_arrow,left,tension=0.6,label=\small{$\omega_\alpha (k)$}}{v1,v2}
\fmf{dbl_plain,right,tension=0.6,label=\small{$y(q)$}}{v1,v2}
\end{fmfgraph*}

\begin{fmfgraph*}(100,36)
\fmfkeep{bubble-fermi}
\fmfleft{in}
\fmfright{out}
\fmfdot{v1}
\fmfdot{v2}
\fmf{dashes_arrow,label=\small{$\omega_\alpha(p)$}}{in,v1}
\fmf{dashes_arrow}{v2,out}
\fmf{plain_arrow,left,tension=0.6,label=\small{$\psi^a (k)$}}{v1,v2}
\fmf{dbl_plain,right,tension=0.6,label=\small{$s^a_\alpha(q)$}}{v1,v2}
\end{fmfgraph*}

\begin{fmfgraph*}(100,36)
\fmfkeep{bubble-yy}
\fmfset{dash_len}{6pt} % this seems to be a local change
\fmfleft{in}
\fmfright{out}
\fmfdot{v1}
\fmfdot{v2}
\fmf{dbl_dashes,label=\small{$ y(p) $}}{in,v1}
\fmf{dbl_dashes}{v2,out}
\fmf{plain_arrow,left,tension=0.6,label=\small{$\omega_\alpha (k)$}}{v1,v2}
\fmf{plain_arrow,left,tension=0.6,label=\small{$\omega_\alpha (q)$}}{v2,v1}
\end{fmfgraph*}

\begin{fmfgraph*}(100,36)
\fmfkeep{bubble-zz}
\fmfset{dash_len}{6pt} % this seems to be a local change
\fmfleft{in}
\fmfright{out}
\fmfdot{v1}
\fmfdot{v2}
\fmf{dbl_dashes,label=\small{$ z_i(p) $}}{in,v1}
\fmf{dbl_dashes}{v2,out}
\fmf{plain_arrow,left,tension=0.6,label=\small{$\psi^a (k)$}}{v1,v2}
\fmf{plain_arrow,left,tension=0.6,label=\small{$\psi^b (q)$}}{v2,v1}
\end{fmfgraph*}

%% psi

\begin{fmfgraph*}(100,36)
\fmfkeep{psi-bubble-1}
\fmfleft{in}
\fmfright{out}
\fmfdot{v1}
\fmfdot{v2}
\fmf{dashes_arrow,label=\small{$\psi^a (p)$}}{in,v1}
\fmf{dashes_arrow}{v2,out}
\fmf{plain_arrow,left,tension=0.6,label=\small{$\omega_\alpha (k)$}}{v1,v2}
\fmf{dbl_plain,right,tension=0.6,label=\small{$s_\alpha^a (q)$}}{v1,v2}
\end{fmfgraph*}

\begin{fmfgraph*}(100,36)
\fmfkeep{psi-bubble-2}
\fmfleft{in}
\fmfright{out}
\fmfdot{v1}
\fmfdot{v2}
\fmf{dashes_arrow,label=\small{$\psi^a (p)$}}{in,v1}
\fmf{dashes_arrow}{v2,out}
\fmf{plain_arrow,left,tension=0.6,label=\small{$\psi^b (k)$}}{v1,v2}
\fmf{dbl_plain,right,tension=0.6,label=\small{$Z^a_b (q)$}}{v1,v2}
\end{fmfgraph*}

%%%%%%%%%%%%%%%%%%%%
%% UNLABLED FOR INTRODUCTION  
%% made a little smaller 22 may

\begin{fmfgraph*}(90,36)
\fmfkeep{bubble-unlab}
\fmfleft{in}
\fmfright{out}
\fmfdot{v1}
\fmfdot{v2}
\fmf{dashes_arrow}{in,v1}
\fmf{dashes_arrow}{v2,out}
\fmf{plain_arrow,left,tension=0.6}{v1,v2}
\fmf{dbl_plain,right,tension=0.6}{v1,v2}
\end{fmfgraph*}

\begin{fmfgraph*}(60,25)
\fmfkeep{tadpole-unlab}
\fmfleft{in,p1}
\fmfright{out,p2}
\fmfdot{c}
\fmf{dashes_arrow}{in,c}
\fmf{dashes_arrow}{c,out}
\fmf{plain_arrow,right, tension=0.8}{c,c}
\fmf{phantom, tension=0.2}{p1,p2}
\end{fmfgraph*}

\begin{fmfgraph*}(60,25)
\fmfkeep{tadpole-heavy-unlab}
\fmfleft{in,p1}
\fmfright{out,p2}
\fmfdot{c}
\fmf{dashes_arrow}{in,c}
\fmf{dashes_arrow}{c,out}
\fmf{dbl_plain,right, tension=0.8}{c,c}
\fmf{phantom, tension=0.2}{p1,p2}
\end{fmfgraph*}

%%%%%%%%%%%%%%%
%% THE END 

\end{fmffile}
}

%% end of my diagrams.tex

\tableofcontents{}

\section{Introduction}

Closed strings in flat spacetime have the property that left- and
right-moving excitations decouple, but this is no longer true in curved
backgrounds such as $AdS_{5}\times S^{5}$ or $AdS_{4}\times CP^{3}$.
Even after one has taken the BMN, or Penrose, limit, focusing on strings
very near to a lightlike trajectory \cite{Berenstein:2002jq}, non-trivial
couplings between the left-movers and right-movers remain. The limit
proposed by Maldacena and Swanson in \cite{Maldacena:2006rv} can
be viewed as a partial restoration of this decoupling, by virtue of
the fact that one light-like worldsheet momentum is taken to be much
larger than the other, $p_{-}\gg p_{+}$.

Both of these limits are taken using a power of the 't Hooft coupling
$\lambda$ as the parameter, thus tying them to the semiclassical
limit $\lambda\gg1$. The BMN limit looks at perturbations of order
$1/\lambda^{1/4}$ from a null geodesic, allowing us (at a given loop
order) to truncate the expansion of the Lagrangian in the number of
fields. The near-flat-space limit then scales the momenta such that
$p_{-}/p_{+}\sim\sqrt{\lambda}$. This can be regarded as a worldsheet
boost in the $\sigma^{-}$ direction, and the simplification comes
from discarding all interaction terms except the leading ones under
this boost. It is not obvious that this truncation will be respected
quantum mechanically. For this to happen, the contributions from where
momenta on internal lines are \emph{not} large must must cancel out,
or at least be sufficiently small.

In the case of strings in $AdS_{5}\times S^{5}$, this quantum consistency
was checked at one loop in \cite{Klose:2007wq} and at two loops in
\cite{Klose:2007rz}. These papers computed primarily corrections
to four-point functions, and compared these to the worldsheet S-matrix
known exactly from integrability \cite{Ahn:2010ka}. The second paper
\cite{Klose:2007rz} also computed corrections to the two-point function,
and compared these to the dispersion relation $E(p)$. It is useful
here to think of the near-flat-space limit as being intermediate between
the BMN limit and the full sigma-model: \begin{equation}
\begin{array}{ll}
p\sim1/\sqrt{\lambda}\:, & \mbox{BMN limit}\\
p\sim\lambda^{-1/4}, & \mbox{near-flat-space}\\
p\sim1, & \mbox{giant magnons.}\end{array}\label{eq:three-p-limits}\end{equation}
We write `giant magnons' for the third sector, as these are the classical
string solutions obeying the dispersion relation at finite $p$ \cite{Hofman:2006xt}.
They are of size $\Delta\phi=p$ along an equator of $S^{5}$.

For the case of strings in $AdS_{4}\times CP^{3}$, relevant for the
comparison with ABJM's superconformal Chern--Simons theory \cite{Aharony:2008ug,Klose:2010ki},
only the bosonic part of the the near-flat-space limit has been studied
\cite{Kreuzer:2008vd}. Fermions are of course essential for one-loop
calculations, and so we work out the complete near-flat-space Lagrangian,
starting from the BMN case of \cite{Sundin:2009zu}. We use this to
calculate corrections to two-point functions, which we can compare
to the exact dispersion relation \begin{equation}
E=\sqrt{\frac{1}{4}+4\, h(\lambda)^{2}\sin^{2}\frac{p}{2}}.\label{eq:dispersion-relation-light}\end{equation}
The interpolating function $h(\lambda)$ is the only part of this
relation not fixed by symmetries. At small values of $\lambda$ it
is $h(\lambda)=\lambda+\mathcal{O}(\lambda^{3})$, but at large $\lambda$
(relevant for semiclassical strings) it is instead \begin{equation}
h(\lambda)=\sqrt{\frac{\lambda}{2}}+c+\mathcal{O}\Big(\frac{1}{\sqrt{\lambda}}\Big).\label{eq:expansion-of-h}\end{equation}
The subleading term $c$ has been the subject of some debate in the
literature. In one-loop energy corrections to spinning strings \cite{McLoughlin:2008ms,Alday:2008ut,Krishnan:2008zs,Gromov:2008fy}
and to giant magnons \cite{Abbott:2010yb}, there is a choice of how
to regulate divergent sums over mode numbers, and using the same momentum
cutoff $\Lambda$ for all modes leads to $c=-\log2/2\pi$. However
it has been argued that it is natural to use a cutoff of $2\Lambda$
for half of the modes (the heavy modes) and this `new' prescription
leads instead to $c=0$ \cite{Gromov:2008fy,Shenderovich:2008bs}.
The same two prescriptions have been implemented in a recent Hamiltonian
analysis \cite{Astolfi:2011ju}, which argues in favour of the new
prescription. Our paper is in some ways the Lagrangian analysis complementary
to this. 

The sigma model in $AdS_{4}\times CP^{3}$ has several novel features
compared to that in $AdS_{5}\times S^{5}$. One is that its excitations
do not all have the same mass: half are `light' ($m=\frac{1}{2}$
in our conventions) and half are heavy ($m=1$). Another is that the
Lagrangian has interactions starting at cubic order, rather than at
quartic order \cite{Astolfi:2008ji,Kreuzer:2008vd,Sundin:2008vt},
greatly expanding the number of Feynman diagrams possible. The cubic
interactions always couple two light modes to one heavy mode, and
therefore lead to bubble diagrams such as\[
\parbox[top][0.45in][c]{1.5in}{\fmfreuse{bubble-unlab}}\]
where we draw the heavy mode as a double line. In this diagram the
same loop momentum applies to both the heavy and the light mode, at
least in the UV. This points towards the `old' prescription, of integrating
up to the same momentum $\Lambda$ for all modes. In this paper we
use primarily dimensional regularisation, but for the simplest case
can also obtain exactly the same results using a momentum cutoff. 

In addition to the term coming from $c=-\log2/2\pi$, we also find
(using either regularisation procedure) some additional terms. These
terms are present in both the BMN and near-flat-space limit, and so
do not represent a problem with the consistency of the near-flat-space
truncation. These extra terms differ for the various particles, vanishing
in the case of corrections to the light fermion propagator. This implies
that they break supersymmetry, possibly as a consequence of dimensional
regularisation. 

In fact these extra terms come (primarily) from the bubble diagram
above, while the term containing $c=-\log2/2\pi$ comes from tadpole
diagrams:\[
\parbox[top][0.4in][c]{0.9in}{\fmfreuse{tadpole-unlab}}+\;\;\parbox{0.9in}{\fmfreuse{tadpole-heavy-unlab}}\]
Each loop here contains only one mode, and so for these tadpole diagrams
there is no obstruction to changing the heavy mode cutoff to implement
the `new' sum. This change cancels the $-\log2/2\pi$ term, and thus
while slightly unsatisfying (since it does nothing to the bubble diagrams)
could be said to lead to $c=0$. 

In this paper we also further the investigation of what one-loop corrections
can teach us about the nature of the heavy modes, as initiated by
Zarembo in \cite{Zarembo:2009au} and continued in \cite{Sundin:2009zu}.
There it was argued that the heavy modes dissolve into a multi-particle
continuum, under the assumption that $c=0$. In our calculation, it
is clear that the decay of a heavy mode into two light modes is kinematically
allowed.

\subsection*{Outline}

In section \ref{sec:The-String-Lagrangian} we set up the theory we
are considering, re-writing some results of \cite{Sundin:2009zu}
for the BMN limit in more convenient notation, and taking the near-flat
space limit of the Lagrangian. 

Sections \ref{sec:Light-Prop}, \ref{sec:A-Detour-through-Cutoffs}
and \ref{sec:Heavy-Prop} study mass corrections to the propagator
in the near-flat space limit. The basic light boson calculation is
quite simple, but the light fermion case is more involved. We discuss
some issues about momentum cutoffs in section \ref{sec:A-Detour-through-Cutoffs}
before working on the heavy modes in section \ref{sec:Heavy-Prop}.

We then turn to the full near-BMN case in section \ref{sec:BMN}.
All of the same issues arise here, but the calculations are a great
deal more complicated and thus less transparent. We can reproduce
our near-flat-space results as limits of these results.

We conclude and summarise in section \ref{sec:Conclusions}. 

Appendix \ref{sec:Fermionic-Matrices} contains extra notation particularly
about fermions. Appendix \ref{sec:Simplifying-the-Quartic-L} contains
manipulations to simplify $\mathcal{L}_{4}$. Appendix \ref{sec:A-DimReg-Options}
looks at expansions of integrals used for dimensional regularisation.

\section{The String Lagrangian and Two Strong-Coupling Limits\label{sec:The-String-Lagrangian}}

Our starting point is the gauge-fixed Lagrangian for type IIA strings
in the near-BMN limit of $AdS_{4}\times CP^{3}$ as derived in \cite{Sundin:2009zu}.
We begin by reviewing this derivation, very briefly.

We are interested strings moving in the quotient super-space \cite{Arutyunov:2008if,Stefanski:2008ik}
\[
\frac{OSP(2,2|6)}{SO(3,1)\times U(3)}\]
the bosonic part of which is $AdS_{4}\times CP^{3}$. The sigma-model
is defined in terms of the following Lie-algebra valued flat current:
\begin{align*}
A_{\mu}(\sigma,\tau) & =-G^{-1}\:\partial_{\mu}G\,,\qquad\qquad G\in OSP(2,2|6)\\
 & =A_{\mu}^{(0)}+A_{\mu}^{(1)}+A_{\mu}^{(2)}+A_{\mu}^{(3)}.\end{align*}
Here $A^{(k)}$ is the component with eigenvalue $i^{k}$ under the
generator of the $\mathbb{Z}_{4}$ automorphism, $\Omega$. The subalgebra
$so(3,1)\oplus u(3)$ is precisely that fixed by $\Omega$, and thus
to be omitted from the action. The $k=0,2$ components are bosonic,
and $k=1,3$ fermionic. The action defined by \cite{Arutyunov:2008if}
is as follows: \begin{equation}
S=\frac{g}{2}\int d^{2}\sigma\:\mbox{Str}\left[\sqrt{-h}h^{\mu\nu}A_{\mu}^{(2)}A_{\nu}^{(2)}+\kappa\:\epsilon^{\mu\nu}A_{\mu}^{(1)}A_{\nu}^{(3)}\right].\label{eq:action-A}\end{equation}
The coupling constant is $g=\sqrt{\lambda/2}=R^{2}/8\pi\alpha'$ which
is taken to be large. In order for local $\kappa$-symmetry to hold
we need $\kappa^{2}=1$, and we now choose $\kappa=1$.%
\footnote{$\kappa$ changes sign under $\sigma\to-\sigma$.%
}

It is necessary to introduce some parameterisation of the group. Since
we are interested in strings near to the null line $\phi=t$, and
will later discard these two directions during gauge fixing, it has
been found convenient to factorise them out from the start. A suitable
parameterisation is \cite{Sundin:2009zu}\[
G=\Lambda(t,\phi)\; F(\chi)\; G_{\perp}(X)\,.\]
The three factors contain the light-cone directions, the fermions
and the eight transverse bosonic directions. Explicitly, \begin{align}
\Lambda(t,\phi) & =\exp\left[\tfrac{i}{2}(x^{+}+(\tfrac{1}{2}-a)x^{-})\Sigma_{+}+\tfrac{i}{4}x^{-}\Sigma_{-}\right]\nonumber \\
G_{\perp}(X) & =G_{AdS}\oplus G_{CP}\;=\;\frac{1+\frac{i}{2}z_{i}\Gamma^{i}}{\sqrt{1-z_{i}^{2}/4}}\oplus\exp\left(W+\bar{W}+\tfrac{1}{2}yT_{5}\right)\label{eq:group-param-factors}\\
F(\chi) & =\chi+\sqrt{1+\chi^{2}}\:.\nonumber \end{align}
Here $z_{i}$ and $t$ are the $AdS{}_{4}$ co-ordinates, while $y$,
$W$, $\bar{W}$ and $\phi$ are the $CP^{3}$ co-ordinates. All the
fermionic fields are contained in the matrix $\chi$, see appendix
A for details. The target-space light-cone coordinates here are non-standard
ones, defined \[
x^{+}=(1-a)t+a\,\phi,\qquad x^{-}=\phi-t,\]
where $a\in[0,1]$ is the same constant as appears in generalised
light-cone gauge.

The bosonic gauge fixing is simplified by introducing an auxiliary
field $\PI$ conjugate to $A^{(2)}$. This brings us to a first-order
formalism in which the Weyl-invariant worldsheet metric $\gamma^{\mu\nu}=\sqrt{-\det h}\, h^{\mu\nu}$
enters as a set of Lagrange multipliers:\begin{equation}
S=-g\int d\sigma\:\mbox{Str}\left[\PI\, A_{0}+\frac{1}{2}\,\epsilon^{\alpha\beta}\, A_{\alpha}^{(1)}\, A_{\beta}^{(3)}-\frac{1}{2\gamma^{00}}\left(\PI^{2}+(A_{1}^{(2)})^{2}\right)+\frac{\gamma^{01}}{\gamma^{00}}\,\PI\, A_{1}^{(2)}\right]\label{eq:action-A-Pi}\end{equation}
By solving for $\PI$ this action is classically equivalent to \eqref{eq:action-A};
for details see \cite{Sundin:2009zu}. The gauge we adopt is the generalised
light-cone gauge of \cite{Arutyunov:2006gs,Klose:2006zd}, in which
we set $x^{+}=\tau$ and the density of the momentum $p_{+}$ to be
a constant: \[
P_{+}=g\int d\sigma\, p_{+}=(1-a)J+a\,\Delta\]
where $\Delta$ and $J$ are the conserved charges from $AdS{}_{4}$
and $CP^{3}$ . Note that this gauge relates the length of the string
worldsheet to $P_{+}$ and the coupling. As a result of this the large
coupling expansion will decompactify the worldsheet. The light-cone
gauge also implies that the physical Hamiltonian is \[
\mathcal{H}_{\mathrm{lc}}=-P_{-}=g\int d\sigma\, p_{-}=J-\Delta.\]
The gauge fixing breaks some of the symmetries and the set of charges
that commutes with the light-cone Hamiltonian combines into $SU(2|2)\times U(1)$
\cite{Bykov:2009jy}.

What we have written thus far is exact in the sense that arbitrarily
large motions of the string are allowed. We now take the BMN, or PP-wave,
limit, in which we focus on strings near to the null geodesic $\phi=t$.
By `near' we mean of order $1/\sqrt{g}$, which means that this limit
is to be taken simultaneously with the semiclassical limit. We implement
this by scaling all fields, making the following replacements in $\mathcal{L}$:
\begin{equation}
x\to\frac{1}{\sqrt{g}}x,\qquad\pi\to\frac{1}{\sqrt{g}}\pi,\qquad\chi\to\frac{1}{\sqrt{g}}\chi\label{eq:scaling-all-fields}\end{equation}
writing $x$ for a generic boson (whose conjugate momentum is $\pi$)
and $\chi$ for a fermion. At large $g$ the Lagrangian is then organised
as an expansion in the number of fields, which we write as follows:\begin{align*}
S & =\int d^{2}\sigma\left[\mathcal{L}_{2}(\pi,x,\chi)+\frac{1}{\sqrt{g}}\mathcal{L}_{3}(\pi,x,\chi)+\frac{1}{g}\mathcal{L}_{4}(\pi,x,\chi)+\mathcal{O}(g^{-3/2})\right].\end{align*}
The result we take from \cite{Sundin:2009zu} as our starting point
is the Lagrangian in this form. (The only change is that we generalise
it to the gauge $a\neq\frac{1}{2}$.) Note that unlike the Hamiltonian
analysis of \cite{Sundin:2009zu,Astolfi:2009qh}, it is not necessary
for us to perform complicated redefinitions of the fermions in order
to make canonically conjugate pairs (up to some order). We can simply
work with them as they stand.

\subsection{The BMN Lagrangian}

The analysis we want to perform in this paper is simplified by eliminating
the momentum variables (introduced for ease of gauge fixing) in favour
of velocities. The equations of motion for the momenta $\pi$ are,
schematically,\[
\frac{\partial\mathcal{L}}{\partial\pi}=0\qquad\Rightarrow\qquad\pi=\partial_{0}x+\frac{1}{\sqrt{g}}\frac{\partial\mathcal{L}_{3}}{\partial\pi}+\frac{1}{g}\frac{\partial\mathcal{L}_{4}}{\partial\pi}+\ldots\]
The higher-order terms on the right mean that the interaction terms
$\mathcal{L}_{3}$ and $\mathcal{L}_{4}$ will change in nontrivial
ways. But the quadratic Lagrangian is essentially%
\footnote{We detail some changes of notation in appendix A.%
} given by replacing $\pi\to\partial_{0}x$ in that of \cite{Sundin:2009zu}:
\begin{align}
\mathcal{L}_{2}= & \frac{1}{2}\partial_{+}y\partial_{-}y+\frac{1}{2}\partial_{+}z_{i}\partial_{-}z_{i}+\frac{1}{4}\partial_{+}\omega_{\alpha}\partial_{-}\bar{\omega}^{\alpha}+\frac{1}{4}\partial_{-}\omega_{\alpha}\partial_{+}\bar{\omega}^{\alpha}-\frac{1}{2}\big(y^{2}+z_{i}^{2}\big)-\frac{1}{8}\omega_{\alpha}\bar{\omega}^{\alpha}\nonumber \\
 & +i\big(\bar{\psi}_{+a}\partial_{-}\psi_{+}^{a}+\bar{\psi}_{-a}\partial_{+}\psi_{-}^{a}\big)+\frac{i}{2}\left[(s_{-})_{\alpha}^{a}\partial_{+}(s_{-})_{a}^{\alpha}+(s_{+})_{\alpha}^{a}\partial_{-}(s_{+})_{a}^{\alpha}\right]\nonumber \\
 & -\frac{1}{2}\big(\bar{\psi}_{-a}\psi_{+}^{a}+\bar{\psi}_{+a}\psi_{-}^{a}\big)-i(s_{+})_{\alpha}^{a}(s_{-})_{a}^{\alpha}.\label{eq:L2-lightcone}\end{align}
Here we have introduced worldsheet light-cone coordinates \[
\sigma^{\pm}=\frac{\tau\pm\sigma}{2},\qquad\Rightarrow\quad\partial_{\pm}=\partial_{0}\pm\partial_{1}.\]
We will refer to both fields and derivatives with a $+$ subscript
as being left-moving, and those with $-$ as right-moving. The list
of fields is as follows: \begin{align}
\psi_{\pm}^{a} & \qquad\mbox{complex light fermion}\qquad a=1,2\nonumber \\
(s_{\pm})_{\alpha}^{a} & \qquad\mbox{real heavy fermion}\qquad\alpha=3,4\label{eq:fields}\\
\omega_{\alpha} & \qquad\mbox{complex light boson}\nonumber \\
y,\; z_{i} & \qquad\mbox{real heavy boson}\qquad i=1,2,3\:.\nonumber \end{align}
The fields transform covariantly under the bosonic part of $SU(2|2)\times U(1)$,
namely $SU(2)^{2}\times U(1)$, and we denote the $SU(2)$ from $AdS_{4}$
with Latin indices and the $SU(2)$ from $CP_{3}$ with Greek. Both
are raised and lowered with $\epsilon$-tensors\[
R^{a}=\epsilon^{ab}R_{b}=\epsilon^{ab}\left(\epsilon_{bc}R^{c}\right),\qquad\epsilon_{12}=\epsilon_{34}=1\quad\Rightarrow\quad\epsilon^{12}=\epsilon^{34}=-1\]
and the action of conjugation is $(C_{a}^{\beta})^{\dagger}=\bar{C}_{\beta}^{a}$.
The light fields also transform under the $U(1)$, with $\psi_{\pm}$
and $\omega_{\alpha}$ having $+1$ charge.

Either by direct inspection or through solving the quadratic equations
of motion, one can derive the following Feynman propagators:\begin{equation}
\left\langle \bar{\omega}^{\alpha}\omega_{\beta}\right\rangle =\delta_{\beta}^{\alpha}\frac{2i}{p^{2}-\frac{1}{4}},\qquad\qquad\left\langle yy\right\rangle =\frac{i}{p^{2}-1},\qquad\qquad\left\langle z_{i}z_{j}\right\rangle =\delta_{ij}\frac{i}{p^{2}-1}.\label{eq:prop-bosonic}\end{equation}
(Notice that our light boson has non-standard normalisation.) For
the fermions: \begin{align}
\left\langle \bar{\psi}_{a}\psi^{b}\right\rangle  & =\delta_{a}^{b}\frac{D_{\psi\psi}}{p^{2}-\frac{1}{4}},\qquad\mbox{where }\; D_{\psi\psi}=\begin{cases}
ip_{+}\quad\mbox{for the case} & \bar{\psi}_{+}\psi_{+}\\
ip_{-} & -,-\\
-i/2 & +,-\mbox{ or }-,+\end{cases}\displaybreak[0]\label{eq:prop-fermi}\\
\left\langle s_{\alpha}^{a}s_{b}^{\beta}\right\rangle  & =\delta_{b}^{a}\delta_{\alpha}^{\beta}\frac{D_{ss}}{p^{2}-1},\qquad D_{ss}=\begin{cases}
ip_{+}\quad\mbox{for } & s_{+}s_{+}\\
ip_{-} & -,-\\
-1 & +,-\\
+1 & -,+\:.\end{cases}\nonumber \end{align}
The cubic Lagrangian is naturally more complicated, and writing $\overleftrightarrow{\partial}_{\pm}=\overleftarrow{\partial}_{\pm}-\overrightarrow{\partial}_{\pm}$
and $Z_{b}^{a}=\sum_{i}z_{i}(\sigma_{i})_{b}^{a}$, it is given by
\begin{align}
\mathcal{L}_{3} & =\frac{i}{8}y\,\big(\omega_{\alpha}\,\overleftrightarrow{\partial_{+}}\bar{\omega}^{\alpha}+\omega_{\alpha}\,\overleftrightarrow{\partial_{-}}\bar{\omega}^{\alpha}\big)+\frac{i}{2}\bar{\psi}_{-a}\,\overleftrightarrow{\partial_{+}}\psi_{-}^{b}\,\partial_{-}Z_{b}^{a}-\frac{i}{2}\bar{\psi}_{+a}\,\overleftrightarrow{\partial_{-}}\psi_{+}^{b}\,\partial_{+}Z_{b}^{a}\nonumber \\
 & +\frac{1}{2}\big(\bar{\psi}_{-a}\partial_{+}\psi_{+}^{b}+\partial_{+}\bar{\psi}_{+a}\psi_{-}^{b}-\bar{\psi}_{+a}\partial_{-}\psi_{-}^{b}-\partial_{-}\bar{\psi}_{-a}\psi_{+}^{b}\big)Z_{b}^{a}\displaybreak[0]\nonumber \\
 & -\epsilon^{ab}\Big[\Big(\frac{3i}{16}(s_{-})_{a}^{\alpha}\,\bar{\psi}_{-b}+\frac{3}{16}(s_{+})_{a}^{\alpha}\,\bar{\psi}_{+b}+\frac{i}{2}(s_{+})_{a}^{\alpha}\,\partial_{-}\bar{\psi}_{-b}-\frac{1}{2}(s_{-})_{a}^{\alpha}\,\partial_{+}\bar{\psi}_{+b}-\frac{1}{4}\partial_{-}(s_{-})_{a}^{\alpha}\,\bar{\psi}_{+b}\nonumber \\
 & -\frac{i}{4}\partial_{-}(s_{-})_{a}^{\alpha}\,\partial_{+}\bar{\psi}_{-b}+\frac{i}{4}\partial_{+}(s_{-})_{a}^{\alpha}\,\partial_{-}\bar{\psi}_{-b}+\frac{i}{4}\partial_{+}(s_{+})_{a}^{\alpha}\,\bar{\psi}_{-b}+\frac{1}{4}\partial_{-}(s_{+})_{a}^{\alpha}\,\partial_{+}\bar{\psi}_{+b}\nonumber \\
 & -\frac{1}{4}\partial_{+}(s_{+})_{a}^{\alpha}\,\partial_{-}\bar{\psi}_{+b}\Big)\omega_{\alpha}-\Big(\frac{i}{8}(s_{-})_{a}^{\alpha}\,\overleftrightarrow{\partial}_{-}\bar{\psi}_{-b}+\frac{1}{8}(s_{+})_{a}^{\alpha}\,\overleftrightarrow{\partial}_{-}\bar{\psi}_{+b}\Big)\partial_{+}\omega_{\alpha}\nonumber \\
 & -\Big(\frac{i}{8}(s_{-})_{a}^{\alpha}\,\overleftrightarrow{\partial}_{+}\bar{\psi}_{-b}+\frac{1}{8}(s_{+})_{a}^{\alpha}\,\overleftrightarrow{\partial}_{+}\bar{\psi}_{+b}\Big)\partial_{-}\omega_{\alpha}\Big]\displaybreak[0]\nonumber \\
 & \negthickspace\negthickspace+\epsilon_{ab}\Big[\frac{3i}{16}(s_{-})_{\alpha}^{a}\,\psi_{-}^{b}-\frac{3}{16}(s_{+})_{\alpha}^{a}\,\psi_{+}^{b}+\frac{i}{2}(s_{+})_{\alpha}^{a}\,\partial_{-}\psi_{-}^{b}+\frac{1}{2}(s_{-})_{\alpha}^{a}\,\partial_{+}\psi_{+}^{b}+\frac{1}{4}\partial_{-}(s_{-})_{\alpha}^{a}\,\psi_{+}^{b}\nonumber \\
 & -\frac{i}{4}\partial_{-}(s_{-})_{\alpha}^{a}\,\partial_{+}\psi_{-}^{b}+\frac{i}{4}\partial_{+}(s_{-})_{\alpha}^{a}\,\partial_{-}\psi_{-}^{b}+\frac{i}{4}\partial_{+}(s_{+})_{\alpha}^{a}\,\psi_{-}^{b}-\frac{1}{4}\partial_{-}(s_{+})_{\alpha}^{a}\,\partial_{+}\psi_{+}^{b}\nonumber \\
 & +\frac{1}{4}\partial_{+}(s_{+})_{\alpha}^{a}\,\partial_{-}\psi_{+}^{b}\Big)\bar{\omega}^{\alpha}-\Big(\frac{i}{8}(s_{-})_{\alpha}^{a}\,\overleftrightarrow{\partial}_{-}\psi_{-}^{b}-\frac{1}{8}(s_{+})_{\alpha}^{a}\,\overleftrightarrow{\partial}_{-}\psi_{+}^{b}\Big)\partial_{+}\bar{\omega}^{\alpha}\nonumber \\
 & -\Big(\frac{i}{8}(s_{-})_{\alpha}^{a}\,\overleftrightarrow{\partial}_{+}\psi_{-}^{b}-\frac{1}{8}(s_{+})_{\alpha}^{a}\,\overleftrightarrow{\partial}_{+}\psi_{+}^{b}\Big)\partial_{-}\bar{\omega}^{\alpha}\Big].\label{eq:L3-BMN}\end{align}
The quartic BMN Lagrangian is very involved and we will not present
it here.

\subsection{The near-flat-space limit\label{sub:The-near-flat-space-limit}}

In the near-flat-space limit we focus on those terms which are important
at large $p_{-}$. We can do this by taking a large worldsheet boost
(in the $\sigma^{-}$, direction) and keeping only the leading terms
under this boost. The quadratic Lagrangian is Lorentz invariant and
so is unaffected, but the cubic and quartic interaction terms break
this symmetry, and are thus simplified in this limit. These simplifications
are the main reason for studying this limit.

The bosonic fields all behave trivially under worldsheet Lorentz transformations.
We have written the fermionic fields in terms of left- and right-moving
components, and it is easy to see that these must scale like $\sqrt{p_{\pm}}$
for $\mathcal{L}_{2}$ to be invariant. Explicitly, the boost involves
the following replacements in $\mathcal{L}$: \begin{equation}
\begin{aligned}\partial_{\pm} & \to g^{\mp1/2}\partial_{\pm}\qquad(\mbox{i.e. }\sigma^{\pm}\to g^{\pm1/2}\sigma^{\pm})\\
\psi_{\pm} & \to g^{\mp1/4}\psi_{\pm}\quad\mbox{and likewise }s_{\pm}\to g^{\mp1/4}s_{\pm}.\end{aligned}
\label{eq:NFS-boost-g}\end{equation}
The leading behaviour of $\mathcal{L}_{3}$ is given by terms that
grow like $\sqrt{g}$, and we keep only these terms. Likewise the
leading terms in $\mathcal{L}_{4}$ grow like $g$. We write the result
of these replacements as follows:\[
\frac{1}{\sqrt{g}}\mathcal{L}_{3}+\frac{1}{g}\mathcal{L}_{4}\;\to\;\mathcal{L}_{3}^{\mathrm{NFS}}+\mathcal{L}_{4}^{\mathrm{NFS}}+\mathcal{O}\Big(\frac{1}{\sqrt{g}}\Big)\]
and after discarding the $\mathcal{O}(1/\sqrt{g})$ terms are thus
studying \begin{equation}
S=\int d^{2}\sigma\left[\mathcal{L}_{2}+\mathcal{L}_{3}^{\mathrm{NFS}}+\mathcal{L}_{4}^{\mathrm{NFS}}\vphantom{\frac{a}{a}}\right].\label{eq:S-NFS}\end{equation}
As noted by \cite{Maldacena:2006rv}, this action has no parameters
at all.

The idea of \cite{Klose:2007wq} is to perform a second boost of the
same form:\begin{equation}
\partial_{\pm}\to\gamma^{\mp1/2}\partial_{\pm},\qquad\psi_{\pm}\to\gamma^{\mp1/4}\psi_{\pm},\qquad s_{\pm}\to\gamma^{\mp1/4}s_{\pm}\label{eq:NFS-boost-gamma}\end{equation}
leading to\begin{equation}
S=\int d^{2}\sigma\left[\mathcal{L}_{2}+\sqrt{\gamma}\mathcal{L}_{3}^{\mathrm{NFS}}+\gamma\mathcal{L}_{4}^{\mathrm{NFS}}\vphantom{\frac{a}{a}}\right].\label{eq:S-NFS-gamma}\end{equation}
If $\gamma$ is kept arbitrary it can be viewed as just a parameter
to keep track of the orders. But if $\gamma=1/g$ (as we will assume)
then this second transformation is the inverse of the first, except
for the fact that we do not recover the interaction terms we discarded.
So what has happened is that we are back in the original variables,
but have baked in the assumption that $p_{-}\gg p_{+}$.

We now write the interaction terms, starting with the cubic term.
Here we simply keep those terms in \eqref{eq:L3-BMN} which grow as
$\sqrt{g}$ under the boost \eqref{eq:NFS-boost-g} (i.e. grow as
fast as the right-moving momentum $p_{-}$): \begin{align}
\mathcal{L}_{3}^{\mathrm{NFS}} & =\frac{i}{8}y\,\omega_{\alpha}\,\overleftrightarrow{\partial_{-}}\bar{\omega}^{\alpha}+\frac{i}{2}\bar{\psi}_{-a}\,\overleftrightarrow{\partial_{+}}\psi_{-}^{b}\,\partial_{-}Z_{b}^{a}-\frac{1}{2}\big(\bar{\psi}_{+a}\partial_{-}\psi_{-}^{b}+\partial_{-}\bar{\psi}_{-a}\psi_{+}^{b}\big)Z_{b}^{a}\nonumber \\
 & -\epsilon^{ab}\Big[\Big(\frac{3i}{16}(s_{-})_{a}^{\alpha}\,\bar{\psi}_{-b}+\frac{i}{2}(s_{+})_{a}^{\alpha}\,\partial_{-}\bar{\psi}_{-b}-\frac{1}{4}\partial_{-}(s_{-})_{a}^{\alpha}\,\bar{\psi}_{+b}-\frac{i}{4}\partial_{-}(s_{-})_{a}^{\alpha}\,\partial_{+}\bar{\psi}_{-b}\nonumber \\
 & \quad\qquad+\frac{i}{4}\partial_{+}(s_{-})_{a}^{\alpha}\,\partial_{-}\bar{\psi}_{-b}\Big)\omega_{\alpha}-\frac{i}{8}(s_{-})_{a}^{\alpha}\overleftrightarrow{\partial_{-}}\,\bar{\psi}_{-b}\partial_{+}\omega_{\alpha}-\frac{i}{8}(s_{-})_{a}^{\alpha}\overleftrightarrow{\partial_{+}}\,\bar{\psi}_{-b}\partial_{-}\omega_{\alpha}\Big]\displaybreak[0]\nonumber \\
 & +\epsilon_{ab}\Big[\Big(\frac{3i}{16}(s_{-})_{\alpha}^{a}\,\psi_{-}^{b}+\frac{i}{2}(s_{+})_{\alpha}^{a}\,\partial_{-}\psi_{-}^{b}+\frac{1}{4}\partial_{-}(s_{-})_{\alpha}^{a}\,\psi_{+}^{b}-\frac{i}{4}\partial_{-}(s_{-})_{\alpha}^{a}\,\partial_{+}\psi_{-}^{b}\nonumber \\
 & \quad\qquad+\frac{i}{4}\partial_{+}(s_{-})_{\alpha}^{a}\,\partial_{-}\psi_{-}^{b}\Big)\bar{\omega}^{\alpha}-\frac{i}{8}(s_{-})_{\alpha}^{a}\overleftrightarrow{\partial_{-}}\,\psi_{-}^{b}\partial_{+}\bar{\omega}^{\alpha}-\frac{i}{8}(s_{-})_{\alpha}^{a}\overleftrightarrow{\partial_{+}}\,\psi_{-}^{b}\partial_{-}\bar{\omega}^{\alpha}\Big].\label{eq:L3-NFS}\end{align}
We notice immediately a distinction from the $AdS_{5}\times S^{5}$
case: this leading order interaction term has both left- and right-moving
fields and derivatives, rather than consisting only of right-moving
objects ($\psi_{-}$, $s_{-}$ and $\partial_{-}$) as in \cite{Maldacena:2006rv}.
This is not unexpected from the form of the boost \eqref{eq:NFS-boost-g}:
in order for a term with two right-moving fermions to scale as $g^{1/2}$,
it must have either no derivatives, or both $\partial_{+}$ and $\partial_{-}$.
Alternatively, it can contain one $\partial_{-}$ with one left-moving
and one right-moving fermion. 

This difference from $AdS_{5}\times S^{5}$ will have an important
consequence. When we come to drawing Feynman diagrams in the next
section, right-moving fields and derivatives will contribute powers
of right-moving momenta $p_{-}$, $k_{-}$ to the numerator, see \eqref{eq:prop-fermi}.
But the presence of left-moving objects will introduce also $k_{+}$,
and a factor $k_{+}k_{-}=k^{2}$ will make any loop integral more
divergent. Because of this, we will have many quadratically divergent
integrals to deal with, while in the $AdS_{5}\times S^{5}$ case \cite{Klose:2007rz},
all of the analogous integrals were finite.

For the quadratic terms, we write $\mathcal{L}_{4}^{\mathrm{NFS}}=\mathcal{L}_{BB}+\mathcal{L}_{BF}+\mathcal{L}_{FF}$,
and the all-boson term is the simplest: \begin{equation}
\mathcal{L}_{BB}=-\frac{1}{8}\Big[\partial_{-}z_{i}\partial_{-}z_{i}+(\partial_{-}y)^{2}+\partial_{-}\bar{\omega}^{\alpha}\partial_{-}\omega_{\alpha}\Big]\Big(z_{j}z_{j}-y^{2}-\frac{1}{4}\bar{\omega}^{\beta}\omega_{\beta}\Big).\label{eq:L4BB-NFS}\end{equation}
For the term mixing bosons and fermions (writing $\bar{\psi}\,\psi=\bar{\psi}_{a}\psi^{a}$
and $\bar{\omega}\,\omega=\bar{\omega}^{\alpha}\omega_{\alpha}$)
we have:\begin{align}
\mathcal{L}_{BF} & =-\frac{i}{8}(s_{-})_{a}^{\alpha}\partial_{-}(s_{-})_{\alpha}^{a}\big(z_{i}^{2}-y^{2}-\frac{1}{2}\omega\,\bar{\omega}\big)+\frac{i}{16}(s_{-})_{a}^{\alpha}(s_{-})_{\gamma}^{a}\,\omega_{\alpha}\overleftrightarrow{\partial_{-}}\,\bar{\omega}^{\gamma}-\frac{i}{32}\bar{\psi}_{-}\overleftrightarrow{\partial_{-}}\,\psi_{-}\,\omega\,\bar{\omega}\nonumber \\
 & +\frac{i}{8}\Big(\bar{\psi}_{-}\,\psi_{-}\,\partial_{-}\omega\,\bar{\omega}+\bar{\psi}_{-}\,\partial_{-}\psi_{-}\,\omega\,\bar{\omega}-\bar{\psi}_{-}\,\psi_{-}\,\omega\,\partial_{-}\bar{\omega}-\partial_{-}\bar{\psi}_{-}\,\psi_{-}\,\omega\,\bar{\omega}\Big)-\frac{i}{8}\bar{\psi}_{-a}\psi_{-}^{b}Z_{c}^{a}\overleftrightarrow{\partial_{-}}Z_{b}^{c}\nonumber \\
 & +\frac{1}{8}y\Big[(s_{-})_{a\alpha}\psi_{-}^{a}\partial_{-}\bar{\omega}^{\alpha}+(s_{-})^{a\alpha}\bar{\psi}_{-a}\partial_{-}\omega_{\alpha}\Big]-\frac{1}{16}\partial_{-}y\Big[(s_{-})_{a\alpha}\psi_{-}^{a}\bar{\omega}^{\alpha}+(s_{-})^{a\alpha}\bar{\psi}_{-a}\omega_{\alpha}\Big]\displaybreak[0]\nonumber \\
 & +\frac{1}{4}y(s_{-})_{a\alpha}\partial_{-}(s_{-})_{c}^{\alpha}Z^{ca}+\frac{i}{4}(s_{-})_{a}^{\alpha}\partial_{-}\bar{\psi}_{-c}\omega_{\alpha}Z^{ca}+\frac{i}{4}(s_{-})_{a\alpha}\partial_{-}\psi_{-}^{b}\,\bar{\omega}^{\alpha}Z_{b}^{a}\nonumber \\
 & +\frac{i}{8}\partial_{-}(s_{-})_{a}^{\alpha}\bar{\psi}_{-c}\omega_{\alpha}Z^{ca}+\frac{i}{8}\partial_{-}(s_{-})_{a\alpha}\psi_{-}^{b}\,\bar{\omega}^{\alpha}Z_{b}^{a}-\frac{i}{8}y^{2}\bar{\psi}_{-}\overleftrightarrow{\partial_{-}}\psi_{-}+\frac{i}{8}(s_{-})_{a\alpha}(s_{-})_{b}^{\alpha}Z_{d}^{a}Z'^{db}\nonumber \\
 & -\frac{1}{8}\Big(i\bar{\psi}_{-}\overleftrightarrow{\partial_{+}}\psi_{-}+\bar{\psi}_{+}\psi_{-}+\bar{\psi}_{-}\psi_{+}\Big)\Big[(\partial_{-}y)^{2}+(\partial_{-}z_{i})^{2}+\partial_{-}\omega_{\alpha}\partial_{-}\bar{\omega}^{\alpha}\Big]\label{eq:L4BF-NFS}\end{align}
At this order, the powers of $g$ under the boost \eqref{eq:NFS-boost-g}
are the same as in the $AdS_{5}\times S^{5}$ case \cite{Maldacena:2006rv}:
terms with two right-moving fermions and $\partial_{-}$ scale correctly
as $g$. On the last line we do also have some terms with left-moving
objects. 

It is possible to remove the terms with left-moving factors from $\mathcal{L}_{BF}$
by nonlinear redefinitions of the fermions, in a way which does not
change final results. Similar terms were sometimes present in the
$AdS_{5}\times S^{5}$ case, in certain parameterisations of the coset.%
\footnote{This is presented in \cite{Sundin2010thesis}. For similar redefinitions,
see \cite{Frolov:2006cc,Arutyunov:2009ga}.%
} We write the last term $\mathcal{L}_{FF}$ in a form obtained by
performing such a redefinition, since the result is much more compact:\begin{align}
\negthickspace\negthickspace\mathcal{L}_{FF} & =\frac{3}{8}\big(\bar{\psi}_{-}\psi_{-}\big)^{2}-\frac{1}{4}\bar{\psi}_{-}\psi_{-}\big(\partial_{+}{\bar{\psi}}_{-}\partial_{-}\psi_{-}+\partial_{-}\bar{\psi}_{-}\partial_{+}{\psi}_{-}\big)-\frac{3}{8}\partial_{+}{\big(\bar{\psi}_{-}\psi_{-}\big)}\partial_{-}\big(\bar{\psi}_{-}\psi_{-}\big)\nonumber \\
 & \negthickspace\negthickspace-\frac{1}{24}(s_{-})_{a}^{\beta}(s_{-})_{\beta}^{d}(s_{-})_{d}^{\gamma}(s_{-})_{\gamma}^{a}-\frac{1}{32}(s_{-})_{\alpha}^{a}(s_{-})_{c}^{\alpha}\,\bar{\psi}_{-a}\,\psi_{-}^{c}+\frac{1}{8}\:\Big[\;\partial_{+}({s}_{-})_{\alpha}^{a}(s_{-})_{c}^{\alpha}\,\bar{\psi}_{-a}\,\partial_{-}\psi_{-}^{c}\nonumber \\
 & \negthickspace\negthickspace\negthickspace\negthickspace+\frac{1}{2}\partial_{+}({s}_{-})_{\alpha}^{a}(s_{-})_{c}^{\alpha}\,\partial_{-}\bar{\psi}_{-a}\,\psi_{-}^{c}+\frac{1}{2}\partial_{+}({s}_{-})_{\alpha}^{a}\partial_{-}(s_{-})_{c}^{\alpha}\,\bar{\psi}_{-a}\,\psi_{-}^{c}+\frac{1}{2}(s_{-})_{\alpha}^{a}\partial_{+}({s}_{-})_{c}^{\alpha}\,\bar{\psi}_{-a}\,\partial_{-}\psi_{-}^{c}\displaybreak[0]\nonumber \\
 & \negthickspace\negthickspace\negthickspace\negthickspace\negthickspace\negthickspace+(s_{-})_{\alpha}^{a}\partial_{+}({s}_{-})_{c}^{\alpha}\,\partial_{-}\bar{\psi}_{-a}\,\psi_{-}^{c}+\frac{1}{2}\partial_{-}(s_{-})_{\alpha}^{a}\partial_{+}({s}_{-})_{c}^{\alpha}\,\bar{\psi}_{-a}\,\psi_{-}^{c}+\frac{1}{2}(s_{-})_{\alpha}^{a}(s_{-})_{c}^{\alpha}\,\partial_{-}\bar{\psi}_{-a}\,\partial_{+}{\psi}_{-}^{c}\nonumber \\
 & \negthickspace\negthickspace\negthickspace\negthickspace\negthickspace\negthickspace\negthickspace\negthickspace+\frac{1}{2}(s_{-})_{\alpha}^{a}(s_{-})_{c}^{\alpha}\,\partial_{+}{\bar{\psi}}_{-a}\,\partial_{-}\psi_{-}^{c}+\frac{3}{2}(s_{-})_{\alpha}^{a}\partial_{-}(s_{-})_{c}^{\alpha}\,\bar{\psi}_{-a}\,\partial_{+}{\psi}_{-}^{c}+\frac{3}{2}\partial_{-}(s_{-})_{\alpha}^{a}(s_{-})_{c}^{\alpha}\,\partial_{+}{\bar{\psi}}_{-a}\,\psi_{-}^{c}\:\Big]\label{eq:L4FF-NFS-final}\end{align}
We discuss details of the redefinitions leading to this form in appendix
\ref{sec:Simplifying-the-Quartic-L}. Note however that this is only
possible for the highest-order interactions. A similar procedure for
the cubic interaction $\mathcal{L}_{3}$ is not allowed. 

In the next three sections we will use the above Lagrangian to calculate
two-point functions on the worldsheet. We return to the full BMN theory
in section \ref{sec:BMN}.

\section{Corrections to the Light Propagators\label{sec:Light-Prop}}

For one light mode, the exact dispersion relation is \[
E=\sqrt{\frac{1}{4}+4\, h(\lambda)^{2}\sin^{2}\frac{\pchain}{2}}\tag{\ref{eq:dispersion-relation-light}}\]
where the interpolating function is $h(\lambda)=\sqrt{\lambda/2}+c+\mathcal{O}(1/\sqrt{\lambda})$.
The momentum here is the one canonically normalised for the spin chain,
and it is this which scales as $\pchain\sim\lambda^{-1/4}$ in the
near-flat-space limit, \eqref{eq:three-p-limits}. Taking this into
account, we can expand $E^{2}$ in $\lambda$ as follows:\begin{align*}
E^{2} & =\frac{\lambda\,\pchain^{2}}{2}+\left[\frac{1}{4}+\sqrt{2\lambda}\, c\,\pchain^{2}-\frac{\lambda\,\pchain^{4}}{24}\right]+\mathcal{O}\Big(\frac{1}{\sqrt{\lambda}}\Big)\\
 & =p_{1}^{2}+\frac{1}{4}+\left(\frac{4cp_{1}^{2}}{\sqrt{2\lambda}}-\frac{p_{1}^{4}}{6\lambda}\right)+\mathcal{O}\Big(\frac{1}{\sqrt{\lambda}}\Big),\qquad\mbox{where }p_{1}=\sqrt{\frac{\lambda}{2}}p_{\mathrm{chain}}\sim\lambda^{1/4}.\end{align*}
On the second line we write it in terms of $p_{1}$ normalised to
match $p_{0}=E$. This $p_{1}$ is the worldsheet momentum in our
uniform lightcone gauge, and its relationship to $\pchain$ could
in principle be deduced from the gauge but we are content just to
read it off here. 

In terms of the worldsheet momenta, the near-flat-space limit assumes
that $p_{-}\gg p_{+}$ (or more strictly $p_{-}\sim\lambda^{1/4}$
and $p_{+}\sim\lambda^{-1/4}$) and thus we can write $(p_{1})^{2}=\frac{1}{4}(p_{+}-p_{-})^{2}=\frac{1}{4}p_{-}+\mathcal{O}(1)$.
Doing this in the correction terms, we arrive at: \begin{align}
p_{0}^{2}-p_{1}^{2} & =\frac{1}{4}+\left(\frac{c\, p_{-}^{2}}{\sqrt{2\lambda}}-\frac{p_{-}^{4}}{96\lambda}\right)+\mathcal{O}\Big(\frac{1}{\sqrt{\lambda}}\Big).\label{eq:light-disp-rel-expanded}\end{align}
We can now read this as giving a mass correction: $p^{2}=m^{2}+\delta m^{2}$,
with $m^{2}=\frac{1}{4}$, and $\delta m^{2}$ in brackets. The \emph{second}
term there is the same two-loop correction as was calculated by \cite{Klose:2007rz}
in the $AdS_{5}\times S^{5}$ case.%
\footnote{Of course the conventional factors of 2 and $\pi$ in \eqref{eq:dispersion-relation-light}
are different. %
} We aim to compute the \emph{first} term of the correction, containing
$c$. The result of \cite{McLoughlin:2008he,Abbott:2010yb} which
we would like to match is that \[
c=-\frac{\log2}{2\pi}.\]
Note that in the expansion \eqref{eq:light-disp-rel-expanded} the
mass squared and its one- and two-loop corrections are all of the
same order in $1/\sqrt{\lambda}$. This is a peculiarity of the near-flat-space
limit. In the BMN case, $p_{\mathrm{chain}}\sim\lambda^{-1/2}$ and
thus $p_{1}\sim\lambda^{0}$, restoring the natural-looking separation
of orders. 

Our calculation of the mass shift comes from writing a familiar geometric
series involving amputated diagrams we call $\mathcal{A}$:%
\footnote{Recall that the interaction terms in \eqref{eq:S-NFS-gamma} are $\sqrt{\gamma}\mathcal{L}_{3}+\gamma\mathcal{L}_{4}$
with $\gamma=1/g=\sqrt{2/\lambda}$.%
}\begin{align}
G_{\omega}(p) & =\frac{2i}{p^{2}-\frac{1}{4}}\sum_{n=0}^{\infty}\left[\gamma\mathcal{A}\frac{2i}{p^{2}-\frac{1}{4}}\right]^{n}\nonumber \\
 & =\frac{2i}{p^{2}-(\frac{1}{4}+\delta m^{2})},\qquad\mbox{ with }\quad\delta m^{2}=2i\gamma\mathcal{A}.\label{eq:light-geom-series}\end{align}
 We will consider corrections not only to the light boson propagator
$\left\langle \bar{\omega}\omega\right\rangle $ as written here,
but also (in section \ref{sub:light-fermi}) to all four of the light
fermion propagators \eqref{eq:prop-fermi}. When correcting for instance
$\left\langle \bar{\psi}_{+}\psi_{+}\right\rangle $, the lines we
amputate will no longer have fixed numerator $2i$ but instead the
various $D_{+s}$ and $D_{r-}$, so (normalising to the bosonic case)
we absorb all of these into $\mathcal{A}$.

There are two topologies of diagrams we must compute, both of which
contain divergences, and it is simplest to employ dimensional regularisation.
We show quite a lot of detail partly in order to attach later discussions
about other possible cutoffs, in section \ref{sec:A-Detour-through-Cutoffs}.

\subsection{Bubble diagrams for $\left\langle \bar{\omega}\omega\right\rangle $}

The loop in each bubble diagram (for a light mode) always contains
one light and one heavy mode. We begin with the simplest case, in
which these are both bosons:\begin{align*}
\mathcal{A}_{B} & =\parbox[top][0.8in][c]{1.5in}{\fmfreuse{bubble}},\qquad q_{\mu}=p_{\mu}-k_{\mu}\\
 & =-1\int\frac{d^{2}k}{(2\pi)^{2}}\;\frac{B}{\left(k^{2}-\frac{1}{4}\right)\left(q^{2}-1\right)},\qquad B=-2\left(\frac{p_{-}+k_{-}}{8}\right)^{2}.\end{align*}
The factor $B$ includes both factors from the propagators \eqref{eq:prop-bosonic}
and from the vertex \eqref{eq:L3-NFS}. The decorations $\partial_{\pm}$
pull down $ik_{\pm}$ from the momentum flowing into the vertex on
that line. 

To treat this integral, we introduce a Feynman parameter $x$ as follows:\begin{align*}
\mathcal{A}_{B} & =-\int\frac{d^{2}k}{(2\pi)^{2}}\int_{0}^{1}dx\frac{B}{\left[x\left(k^{2}-\frac{1}{4}\right)+(1-x)\left((p-k)^{2}-1\right)\right]^{2}}\\
 & =-\int_{0}^{1}dx\int\frac{d^{2}\ell}{(2\pi)^{2}}\:\frac{B}{\left[\ell^{2}-\Delta(x)\right]^{2}}\end{align*}
where\[
\ell_{\mu}=k_{\mu}-(1-x)p_{\mu},\qquad\qquad\Delta(x)=\frac{1}{4}x^{2}-x+1\,.\]
This effective mass squared $\Delta(x)$ assumes $p^{2}=\frac{1}{4}$,
i.e. that we are very near to the pole of the original propagator.
In terms of $\ell_{\mu}$, \[
B=-\frac{(x-2)^{2}}{32}p_{-}^{2}+\frac{2(x-2)\ell_{-}p_{-}-\ell_{-}^{2}}{32}.\]
Next we rotate the $\ell_{0}$ contour to the imaginary axis, and
in the resulting Euclidean space, the terms $\ell_{-}^{n}$ will vanish.%
\footnote{We define $\ell_{0}=iL_{0}$, $\ell_{1}=L_{1}$ and $L_{\mu}=L\,(\sin\theta,\cos\theta)$.
Then the denominator is $[L^{2}-\Delta(x)]^{2}$, and in the numerator,
any nonzero power $\ell_{-}^{n}$ will lead to $\int_{0}^{2\pi}d\theta\: e^{-in\theta}=0$.%
} The radial integral then gives $1/2\Delta(x)$, and finally we obtain:\begin{align}
\mathcal{A}_{B} & =\frac{i}{32\pi}p_{-}^{2}\qquad\qquad\Rightarrow\:\delta m^{2}=2i\gamma\mathcal{A}_{B}=\frac{-1}{16\pi}\gamma p_{-}^{2}.\label{eq:ww-B-bubble}\end{align}

Next we treat the case where the heavy and light modes in the loop
are both fermions:\begin{align}
\mathcal{A}_{F}(p) & =\parbox[top][0.8in][c]{1.5in}{\fmfreuse{bubble-fermi}}\nonumber \\
 & =2\int\frac{d^{2}k}{(2\pi)^{2}}\frac{B}{(k^{2}-\frac{1}{4})(q^{2}-1)}\displaybreak[0]\label{eq:first-A_F-integral}\\
 & =2\int_{0}^{1}dx\int\frac{d^{2}\ell}{(2\pi)^{2}}\frac{1}{\left[\ell^{2}-\Delta(x)\right]^{2}}\sum_{s,s'=0}^{2}\ell_{+}^{s}\ell_{-}^{s'}B_{ss'}\nonumber \\
 & =2\int_{0}^{1}dx\left[I_{2}^{s=0}B_{00}+I_{2}^{s=1}B_{11}+I_{2}^{s=2}B_{22}\right].\nonumber \end{align}
Here $B$ has the same meaning,%
\footnote{Somewhat arbitrarily, the factors from the fermion loop (i.e. re-ordering
fields) and the $\sum_{a}$ in the loop are part of the prefactor,
not $B$.%
} but is much more complicated since, expanding $(\mathcal{L}_{3})^{2}$,
there are in all 25 terms, and the propagators \eqref{eq:prop-fermi}
give various powers of the momenta. After writing $B$ in terms of
$\ell_{+}^{s}\ell_{-}^{s'}$, terms with $s\neq s'$ still vanish,
but there are now also divergent terms, with one or two powers of
$\ell_{+}\ell_{-}=\ell^{2}$ in the numerator. 

As we noted below \eqref{eq:L3-NFS}, the fact that left-moving momenta
$k_{+}$ and $q_{+}$ appear in $B$, and produce divergences, is
a crucial qualitative difference from the $AdS_{5}\times S^{5}$ case. 

We will treat the resulting divergent integrals using dimensional
regularisation. Here is the generic integral, including the case $n\neq2$
for later use:%
\footnote{Here of course $2\pi^{d/2}/\Gamma(\frac{d}{2})$ is the volume of
a unit $(d\mbox{-}1)$-sphere, and $\gamma_{E}=0.577\ldots$.%
}\begin{align}
I_{n}^{s}(\Delta) & =\int\frac{d^{d}\ell}{(2\pi)^{d}}\,\frac{(\ell_{+}\ell_{-})^{s}}{\left[\ell^{2}-\Delta\right]^{n}}\:,\qquad\qquad\qquad d=2-\epsilon\nonumber \\
 & =\frac{i}{(2\pi)^{d}}\,\frac{2\pi^{d/2}}{\Gamma(\frac{d}{2})}\int_{0}^{\infty}dL\,\frac{(-1)^{s}L^{2s+d-1}}{\left[L^{2}+\Delta\right]^{n}}\displaybreak[0]\nonumber \\
 & =\frac{i(-1)^{s}}{(4\pi)^{d/2}}\,\frac{2}{\Gamma(\frac{d}{2})}\:\left(\frac{1}{\Delta}\right)^{n-s-d/2}\frac{\Gamma(n-s-\frac{d}{2})\Gamma(s+\frac{d}{2})}{2\Gamma(n)}\displaybreak[0]\label{eq:dimreg-general-integral}\\
 & =\frac{i}{2\pi}\begin{cases}
\frac{1}{2\Delta},\qquad\qquad\qquad\qquad\qquad\qquad n=2\quad\mbox{ and } & s=0\\
-\frac{1}{\epsilon}+\frac{\gamma_{E}-\log\pi-2\log2}{2}+\frac{1}{2}+\frac{\log\Delta}{2}\;+\mathcal{O}(\epsilon), & s=1\\
2\Delta\left(-\frac{1}{\epsilon}+\frac{\gamma_{E}-\log\pi-2\log2}{2}+\frac{1}{4}+\frac{\log\Delta}{2}\right)\;+\mathcal{O}(\epsilon), & s=2\,.\end{cases}\nonumber \end{align}
(See appendix \ref{sec:A-DimReg-Options} for comment on these expansions.) 

The coefficients of these integrals are as follows, after using $p_{+}p_{-}=\frac{1}{4}$
: \begin{align*}
B_{00} & =-\frac{3x^{4}-6x^{3}-8x^{2}+20x-8}{32}p_{-}^{2}\displaybreak[0]\\
B_{11} & =-\frac{19x^{2}-22x}{16}p_{-}^{2}\\
B_{22} & =-\frac{1}{4}p_{-}^{2}\:.\end{align*}
After integrating on $x$, the log divergence $s=1$ will cancel against
the quadratic divergence $s=2$. Then the final result for the fermion
bubble is \begin{equation}
\mathcal{A}_{F}=\frac{+i}{16\pi}p_{-}^{2}\qquad\Rightarrow\:\delta m^{2}=\frac{-1}{8\pi}\gamma p_{-}^{2}\:.\label{eq:ww-F-bubble}\end{equation}

\subsection{Tadpole diagrams for $\left\langle \bar{\omega}\omega\right\rangle $}

The tadpoles made with two 3-vertices can be seen to vanish by looking
at $\mathcal{L}_{3}$, equation \eqref{eq:L3-NFS}:\[
\parbox[top][0.6in][c]{1in}{\fmfreuse{lollipop}}=0\,.\]
But the ones constructed with one 4-vertex are more interesting. We
use \eqref{eq:L4BF-NFS} to calculate the first two diagrams. The
only nonzero terms are those where both decorations $\partial_{-}$
act on the external legs, which for $\omega$ running in the loop
gives\begin{align*}
\mathcal{A}_{T1}=\parbox[top][0.6in][c]{1in}{\fmfreuse{tadpole}} & =i\frac{1}{32}\,2\, p_{-}^{2}\int\frac{d^{d=2-\epsilon}k}{(2\pi)^{2}}\frac{2i}{k^{2}-\frac{1}{4}}\\
 & =\frac{-p_{-}^{2}}{8}I_{n=1}^{s=0}(\Delta=\tfrac{1}{4})\,.\end{align*}
For a heavy mode running in the loop, we get: \begin{align}
\mathcal{A}_{T2}=\parbox[top][0.6in][c]{1in}{\fmfreuse{tadpole-heavy}} & =i\frac{-1}{8}\,(3-1)\, p_{-}^{2}\int\frac{d^{2}k}{(2\pi)^{2}}\frac{i}{k^{2}-1}\nonumber \\
 & =\frac{p_{-}^{2}}{4}I_{1}^{0}(1)\,.\label{eq:ww-AT2-tadpole-s}\end{align}

There are also diagrams of this topology with a light fermion running
in the loop. The terms in \eqref{eq:L4BF-NFS} which contribute are
those on the last line, and they give the following two integrals
(including the only quadratically divergent tadpole):\begin{align*}
\mathcal{A}_{T3}=\parbox[top][0.6in][c]{1in}{\fmfreuse{tadpole-fermi}} & =i\frac{-i}{8}\,2\, p_{-}^{2}\int\frac{d^{2}k}{(2\pi)^{2}}\frac{2k_{+}k_{-}}{k^{2}-\frac{1}{4}}+i\frac{-1}{8}4p_{-}^{2}\int\frac{d^{2}k}{(2\pi)^{2}}\frac{-i/2}{k^{2}-\frac{1}{4}}\\
 & =\frac{p_{-}^{2}}{2}I_{1}^{1}(\tfrac{1}{4})-\frac{p_{-}^{2}}{4}I_{1}^{0}(\tfrac{1}{4})\,.\end{align*}
The diagrams with a heavy fermion in the loop can be seen to vanish
by just looking at \eqref{eq:L4BF-NFS}. 

All of the tadpole integrals are $n=1$ cases of \eqref{eq:dimreg-general-integral}
above, with $\Delta=\frac{1}{4}$ or $1$. The expansions needed are:\begin{equation}
I_{n=1}^{s}(\Delta)=\frac{i}{2\pi}\begin{cases}
-\frac{1}{\epsilon}+\frac{\gamma_{E}-\log\pi-2\log2+\log\Delta}{2}+\mathcal{O}(\epsilon), & s=0\\
-\frac{\Delta}{\epsilon}+\Delta\frac{\gamma_{E}-\log\pi-2\log2+\log\Delta}{2}+\mathcal{O}(\epsilon),\qquad & s=1.\end{cases}\label{eq:dimreg-tadpole-expansions}\end{equation}
Adding the bosonic and fermionic terms, the divergences cancel leaving
a total tadpole contribution of \begin{equation}
\mathcal{A}_{T}=\frac{i}{4}p_{-}^{2}\frac{\log2}{2\pi}\qquad\Rightarrow\:\delta m_{T}^{2}=-\frac{1}{2}\gamma p_{-}^{2}\frac{\log2}{2\pi}\,.\label{eq:ww-tadpole}\end{equation}

\subsection{Total for the light boson}

Adding up all of these terms, the total amplitude is $\mathcal{A}_{B}+\mathcal{A}_{F}+\mathcal{A}_{T}=\frac{i}{4}p_{-}^{2}\frac{\log2}{2\pi}+\frac{3i}{32\pi}p_{-}^{2}.$
As a mass correction this is\begin{equation}
\delta m^{2}=-\frac{\log2}{2\pi}\frac{\gamma}{2}p_{-}^{2}-\frac{3}{16\pi}\gamma p_{-}^{2}.\label{eq:ww-total-dm2}\end{equation}
Comparing this to \eqref{eq:light-disp-rel-expanded}, we see that
the term with $\log2$ (from the tadpoles) is perfect so long as $\gamma=\sqrt{2/\lambda}$.
The extra term is perhaps unexpected, and we commend on this in the
conclusions.

\subsection{Corrections for the light fermion $\left\langle \bar{\psi}\psi\right\rangle $\label{sub:light-fermi}}

We now turn to the case of corrections to the light fermion propagators,
\eqref{eq:prop-fermi}. The two bubble diagrams we need are these:\begin{align*}
\mathcal{A}_{\psi} & =\quad\parbox[top][0.7in][c]{1.4in}{\fmfreuse{psi-bubble-1}}+\;\parbox[top][0.6in][c]{1.5in}{\fmfreuse{psi-bubble-2}}\\
 & =\int\frac{d^{2}k}{(2\pi)^{2}}\frac{B}{(k^{2}-\frac{1}{4})(q^{2}-1)}\,.\end{align*}
The integrals are of course the same as \eqref{eq:first-A_F-integral}
above, with different coefficients $B=\sum_{s,s'}\ell_{+}^{s}\ell_{-}^{s'}B_{ss'}$.
As noted above, $B$ now includes the numerators $D_{\psi\psi}$ of
the two amputated propagators (as well as the vertices and internal
propagators). We must calculate the coefficients separately for each
of the four cases $\left\langle \bar{\psi}_{+}\psi_{+}\right\rangle $,
$\left\langle \bar{\psi}_{-}\psi_{-}\right\rangle $, $\left\langle \bar{\psi}_{-}\psi_{-}\right\rangle $
and $\left\langle \bar{\psi}_{+}\psi_{+}\right\rangle $, and the
results are all different until the point at which we use $p^{2}=\frac{1}{4}$
to get the shift near to the pole of the unperturbed propagator. Then
they all agree, and we obtain \begin{align*}
B_{00} & =-\frac{1}{128}\left(64+24x-32x^{2}-8x^{3}+x^{4}+2x^{5}\right)p_{-}^{2}\displaybreak[0]\\
B_{11} & =-\frac{1}{16}\left(-12-7x+x^{2}+6x^{3}\right)p_{-}^{2}\\
B_{22} & =-\frac{1}{8}(6x-1)p_{-}^{2}\:.\end{align*}
Then using the expansions of the integrals in \eqref{eq:dimreg-general-integral},
we find \begin{align}
\delta m^{2}= & \frac{1}{64\pi}\gamma\Big(-\frac{42}{\epsilon}+21\gamma_{E}-26\log2+21\log\pi\Big)p_{-}^{2}\,.\label{eq:psi-bubble}\end{align}
There is an important observation we can make at this stage. In this
expansion in $1/\epsilon$, there is no order $1$ term with a rational
coefficient (inside the bracket).%
\footnote{Such terms do occur in the individual bubble diagrams drawn, which
contribute $\pm\frac{3}{64\pi}p_{-}^{2}$, but these cancel out of
the total.%
} Since no such terms will appear in any of the tadpole integrals \eqref{eq:dimreg-tadpole-expansions},
we conclude that there will be no `extra' term like that seen in \eqref{eq:ww-total-dm2}.
A further difference from the $\omega$ case above is that the total
of the bubble diagrams is not finite.

Next we turn to the tadpole diagrams, using \eqref{eq:L4BF-NFS} and
\eqref{eq:L4FF-NFS-final}. We must compute these for each of the
four cases $\left\langle \bar{\psi}_{+}\psi_{+}\right\rangle $, $\left\langle \bar{\psi}_{-}\psi_{-}\right\rangle $,
$\left\langle \bar{\psi}_{-}\psi_{-}\right\rangle $ and $\left\langle \bar{\psi}_{+}\psi_{+}\right\rangle $
separately, but again the final coefficients work out to be the same.%
\footnote{Note also that whether we use \eqref{eq:L4FF-NFS-initial} or the
simplified \eqref{eq:L4FF-NFS-final} makes no difference to these
coefficients.%
} Those with a light mode in the loop are \begin{align*}
\mathcal{A}_{L} & =\parbox[top][0.8in][c]{1in}{\fmfreuse{psi-tad-w}}+\parbox[top][0.6in][c]{1in}{\fmfreuse{psi-tad-psi}}\\
 & =p_{-}^{2}\left[\frac{9}{16}I_{1}^{0}(\tfrac{1}{4})-\frac{3}{8}I_{1}^{1}(\tfrac{1}{4})\right].\end{align*}
With a heavy mode in the loop, we find \begin{align}
\mathcal{A}_{H} & =\parbox[top][0.8in][c]{1in}{\fmfreuse{psi-tad-yz}}+\parbox[top][0.6in][c]{1in}{\fmfreuse{psi-tad-s}}\nonumber \\
 & =p_{-}^{2}\left[-\frac{3}{2}I_{1}^{0}(1)+\frac{3}{8}I_{1}^{1}(1)\right].\label{eq:psi-heavy-integrals}\end{align}
After using the expansions of the integrals in \eqref{eq:dimreg-tadpole-expansions},
the total is as follows: \begin{align*}
\delta m^{2}=\gamma & \frac{21}{64\pi}\Big(-\frac{2}{\epsilon}+\gamma_{E}+\log2-2\log2\Big)p_{-}^{2}\,.\end{align*}
Happily the divergent $1/\epsilon$ term cancels that of \eqref{eq:psi-bubble}.

Adding the bubble and tadpole contributions gives us that the full
mass shift, which for all the $\left\langle \bar{\psi}\psi\right\rangle $
cases is \begin{equation}
\delta m^{2}=-\gamma\frac{\log2}{4\pi}p_{-}^{2}\label{eq:psi-total-dm2}\end{equation}
This is exactly what we would expect from $c=-\frac{\log2}{2\pi}$
in \eqref{eq:light-disp-rel-expanded}. Unlike the bosonic case \eqref{eq:ww-total-dm2},
there are no extra terms.

\section{A Detour via some Delicate Cutoffs\label{sec:A-Detour-through-Cutoffs}}

In this section we look at some effects of employing explicit energy
or momentum cutoffs $\Lambda$ instead of dimensional regularisation. 

The main reasons for doing so are to try to better understand the
extra terms in the mass correction \eqref{eq:ww-total-dm2}, and to
explore the prescription-dependence of $c$ by looking at `old' and
`new' cutoffs (in section \ref{sub:Old-and-new-prescriptions}). We
will also get extra cross-checks between the bubble ($\mathcal{L}_{3}$)
and the tadpole ($\mathcal{L}_{4}$) calculations. 

Most of this section will be about only the light boson case. We comment
briefly on other cases in section \ref{sub:Modes-other-than-w}.

\subsection{Using hard cutoffs\label{sub:Using-hard-cutoffs}}

The first observation is that if we work with hard momentum cutoffs
$\left|k\right|<\Lambda$, instead of dimensional regularisation,
then the cancellation of infinities is different. Each of $\mathcal{A}_{F}$
and $\mathcal{A}_{T}$ contains a logarithmic and a quadratic divergence,
and above we cancelled these \emph{within} them, since all are $1/\epsilon$
terms when working in $d=2-\epsilon$ dimensions. However in terms
of a hard cutoff, instead these $\Lambda^{2}$ terms cancel \emph{between}
them. This cancellation provides a useful cross-check between the
bubble ($\mathcal{L}_{3}$) and the tadpole ($\mathcal{L}_{4}$) calculations.

The only quadratically divergent terms are the $(\ell_{+}\ell_{-})^{2}$
part of the fermionic bubble integral%
\footnote{We note in passing the following issue \cite{Leibbrandt:1975dj}.
In four dimensions, the shift from integrating over $k_{\mu}$ with
$\left|k\right|<\Lambda$ to integrating over $\ell_{\mu}$ with $\left|\ell\right|<\Lambda$
would change the value of the $\Lambda^{2}$-divergent integral, and
thus not be allowed. (The finite and $\log\Lambda$ integrals are
unchanged.) This is however not a problem in two dimensions. Nor is
it a problem in dimensional regularisation, in any number of dimensions. %
}\begin{align}
\mathcal{A}_{F}^{s=2} & =2B_{22}\frac{i}{2\pi}\int_{0}^{1}dx\int_{0}^{\Lambda}dL\frac{L^{5}}{[L^{2}+\Delta(x)]^{2}}\nonumber \\
 & =\frac{-ip_{-}^{2}}{4\pi}\left[\frac{\Lambda^{2}}{2}-\frac{7}{6}\log\Lambda+\frac{\log2}{6}-\frac{7}{72}+\mathcal{O}\left(\frac{1}{\Lambda}\right)\right]\label{eq:L^2-fermi-bubble}\end{align}
and the fermionic tadpole integral \begin{align}
\mathcal{A}_{T3} & =\frac{p_{-}^{2}}{2}\frac{i}{2\pi}\int_{0}^{\Lambda}dK\frac{K^{3}}{K^{2}+\frac{1}{4}}-\frac{p_{-}^{2}}{4}\frac{i}{2\pi}\int_{0}^{\Lambda}dK\frac{-K}{K^{2}+\frac{1}{4}}\nonumber \\
 & =\frac{ip_{-}^{2}}{4\pi}\left[\frac{\Lambda^{2}}{2}-\frac{1}{4}\log\Lambda-\frac{\log2}{4}\right]+\frac{ip_{-}^{2}}{4\pi}\left[2\log\Lambda+2\log2\vphantom{\frac{1}{1}}\right]+\mathcal{O}\left(\frac{1}{\Lambda}\right).\label{eq:L^2-fermi-tadpole}\end{align}
The $\log\Lambda$ divergences similarly cancel. They are:\begin{align}
\mathcal{A}_{F}^{s=1} & =2\int_{0}^{1}dx\: B_{11}\int_{0}^{\Lambda}dL\frac{-L^{3}}{[L^{2}+\Delta(x)]^{2}}=\frac{ip_{-}^{2}}{4\pi}\left[-\frac{7}{6}\log\Lambda-\frac{17\log2}{6}+\frac{155}{72}+\ldots\right]\nonumber \\
\mathcal{A}_{T1} & =\frac{-p_{-}^{2}}{8}\int_{0}^{\Lambda}dK\frac{-K}{K^{2}+\frac{1}{4}}=\frac{ip_{-}^{2}}{4\pi}\left[\frac{1}{4}\log\Lambda+\frac{1}{4}\log2+\ldots\right]\nonumber \\
\mathcal{A}_{T2} & =\frac{p_{-}^{2}}{4}\frac{i}{2\pi}\int_{0}^{\Lambda}dK\frac{-K}{K^{2}+1}=\frac{ip_{-}^{2}}{4\pi}\left[\frac{-1}{2}\log\Lambda+\ldots\right]\label{eq:log-L-heavy-boson-tadpole}\end{align}
and the finite terms are of course unchanged:\begin{align*}
\mathcal{A}_{F}^{s=0} & =\int_{0}^{1}dx\:2B_{00}\frac{1}{2\Delta(x)}=\frac{ip_{-}^{2}}{4\pi}\left[3\log2-2\right]\\
\mathcal{A}_{B} & =\frac{ip_{-}^{2}}{4\pi}\frac{1}{8}.\end{align*}

Adding these up, we obtain the same finite result as before: \[
\delta m^{2}=-\frac{\log2}{2\pi}\frac{\gamma}{2}p_{-}^{2}-\frac{3}{16\pi}\gamma p_{-}^{2}.\tag{\ref{eq:ww-total-dm2}}\]
Note in particular that we have exactly the same `extra' term as we
did using dimensional regularisation. We now turn to looking for other
explanations of this term, for this case.

\subsection{High-precision cutoffs\label{sub:High-precision-cutoffs}}

Notice that the finite part after the cancellation of quadratic divergences
(between bubble and tadpole) is sensitive to very small changes in
the cutoff. In this section we investigate the effects of such small
changes, replacing $\Lambda$ with one of these: \begin{equation}
\begin{array}{ll}
\Lambda_{\mathrm{light}}=\Lambda-\dfrac{\iota}{2\Lambda}, & \mbox{light modes}\\
\Lambda_{\mathrm{heavy}}=\Lambda-\dfrac{\varkappa\vphantom{1^{2}}}{2\Lambda}, & \mbox{heavy modes.}\end{array}\label{eq:hyperfine-cutoffs}\end{equation}
For the next page or so we leave $\iota$ and $\varkappa$ arbitrary. 
\begin{itemize}
\item For the tadpole we can simply write down the result, since the only
$\Lambda^{2}$ divergent integral \eqref{eq:L^2-fermi-tadpole} involves
only a light mode. The effect of changing from $\Lambda\to\Lambda_{\mathrm{light}}$
is one iota: \begin{equation}
\Lambda^{2}\to\Lambda^{2}-\iota\:.\label{eq:hyperfine-change-tadpole}\end{equation}

\item For the bubble's $s=2$ term \eqref{eq:L^2-fermi-bubble}, we expect
some change involving both heavy and light modes, but it is not obvious
what mixture of $\iota$ and $\kappa$ should appear. In fact it is
not clear that there will be any way of implementing \emph{both} cutoffs
\eqref{eq:hyperfine-cutoffs} in the same loop integral. But we can
make an attempt, by introducing Pauli--Villars regulators \cite{Pauli:1949zm},
replacing in the heavy and light propagators the following:\begin{align*}
\frac{1}{k^{2}-\frac{1}{4}}\quad & \underset{\mathrm{PV}}{\longrightarrow}\quad\frac{1}{k^{2}-\frac{1}{4}}-\frac{1}{k^{2}-\Lambda_{\mathrm{light}}^{2}}\\
\frac{1}{q^{2}-1}\quad & \underset{\mathrm{PV}}{\longrightarrow}\quad\frac{1}{q^{2}-1}-\frac{1}{q^{2}-\Lambda_{\mathrm{heavy}}^{2}}\,.\end{align*}
The propagators are unchanged for $k^{2}\ll\Lambda^{2}$, but when
$k^{2}\gg\Lambda^{2}$ they die like $1/k^{4}$. Defining a generalisation
of the integral in \eqref{eq:first-A_F-integral} to allow two arbitrary
masses\[
J(m,M)=\int\frac{d^{2}k}{(2\pi)^{2}}\frac{B}{\left(k^{2}-m^{2}\right)(q^{2}-M^{2})}\]
the effect of imposing the Pauli--Villars regulators is to replace
$J(\tfrac{1}{2},1)$ with four terms: \[
J(\tfrac{1}{2},1)\quad\underset{\mathrm{PV}}{\longrightarrow}\quad J(\tfrac{1}{2},1)-J(\Lambda_{\mathrm{light}},1)-J(\tfrac{1}{2},\Lambda_{\mathrm{heavy}})+J(\Lambda_{\mathrm{light}},\Lambda_{\mathrm{heavy}})\:.\]
Each of these terms can be treated exactly as we did before,%
\footnote{This means we are regulating each of the four terms with dimensional
regularisation. We can instead use another hard cutoff $\Lambda'$
for all the terms, and as long as $\Lambda'\gg\Lambda_{\mathrm{heavy}},\Lambda_{\mathrm{light}}$
the result is identical. (If $\Lambda'=\Lambda$ then \eqref{eq:hyperfine-change-PV-bubble}
becomes $\Lambda^{2}\to\Lambda^{2}-(\iota+\varkappa)$ instead.)%
} simply re-using the integrals \eqref{eq:dimreg-general-integral}
with the appropriate effective masses $\Delta(x,m,M)$. Thus the $s=2$
part of each is given by this:\[
J^{s=2}(m,M)=\int_{0}^{1}\negthickspace\negthickspace dx\int\frac{d^{2-\epsilon}\ell}{(2\pi)^{2}}\frac{(\ell_{+}\ell_{-})^{2}}{\left[\ell^{2}-\Delta(x,m,M)\right]^{2}},\qquad\negthickspace\negthickspace\begin{aligned}\ell_{\mu} & =k_{\mu}-(1-x)p_{\mu}\\
\Delta & =x\, m^{2}+(1-x)M^{2}+x(1-x)p^{2}.\end{aligned}
\]
Notice that $\ell_{\mu}$ is defined in the same way in all four terms,
so that the identification of the term $(\ell_{+}\ell_{-})^{2}$ in
$B$ is the same here as before. (It comes with the same coefficient
$B_{22}$.) The final result for the $s=2$ integral (at $p^{2}=\frac{1}{4}$)
is then \begin{equation}
J_{\mathrm{PV}}^{s=2}(\tfrac{1}{2},1)=\frac{\Lambda^{2}}{2}-\frac{7}{6}\log\Lambda+\left(\frac{\log2}{6}-\frac{5}{48}-\frac{\iota+\varkappa}{4}\right)+\mathcal{O}\left(\frac{1}{\Lambda}\right).\label{eq:L^2-Int-s2-PV}\end{equation}
Thus the effect on the bubble diagram of turning on $\iota,\varkappa$
can be summarised as \begin{equation}
\Lambda^{2}\to\Lambda^{2}-\frac{\iota+\varkappa}{2}\:.\label{eq:hyperfine-change-PV-bubble}\end{equation}
This makes some sense, in that if you change both cutoffs by the same
amount $\iota=\kappa$ then this becomes an overall change of $\Lambda^{2}$.
What you can't predict without calculation is that the effect of $\iota\neq\varkappa$
shouldn't be some other mixture.
\end{itemize}
The total effect on $\mathcal{A}$ of both \eqref{eq:hyperfine-change-tadpole}
and \eqref{eq:hyperfine-change-PV-bubble} is then as follows. Using
\eqref{eq:L^2-fermi-bubble} and \eqref{eq:L^2-fermi-tadpole} above,
we have $\mathcal{A}\to\mathcal{A}+\frac{ip_{-}^{2}}{8\pi}\left(\iota-\frac{\iota+\varkappa}{2}\right)$,
or, in terms of the mass shift, \[
\delta m^{2}=2i\mathcal{A}\to\delta m^{2}+(\varkappa-\iota)\frac{\gamma p_{-}^{2}}{8\pi}.\]
It remains to argue what tiny shifts $\iota,\varkappa$ we should
use. 

Ideally one would like to impose exactly the same cutoff on the physical
energy of all the modes. Unlike the worldsheet momentum $k_{1}$,
or worse the Euclidean energy $K_{0}$, the energy is a physical,
gauge-invariant quantity. 

One argument is this: in the original Lorentzian momentum, $E^{2}=k_{0}^{2}=k_{1}^{2}+m^{2}$
on the mass shell. Imposing $\left|E\right|<\Lambda$ thus reads $k_{1}^{2}<\Lambda^{2}-m^{2}$.
If we impose this momentum cutoff on both of the Euclidean directions,
then we are led to $K^{2}<\Lambda^{2}-m^{2}$. For the light modes
this means $\iota=\frac{1}{4}$, while for the heavy modes $\varkappa=1$.
This gives\begin{equation}
\delta m^{2}\to\delta m^{2}+\frac{3}{32}\frac{p_{-}^{2}\gamma}{\pi}.\label{eq:hyperfine-change-ww}\end{equation}
Sadly this is too small to cancel the extra term in \eqref{eq:ww-total-dm2},
by a factor of 2. 

The attentive reader will by this point have smelled something a little
fishy. Not only have we made up a fairly strange cutoff, \eqref{eq:hyperfine-cutoffs},
we have also mixed the results of using this in the Pauli--Villars
bubble integral \eqref{eq:L^2-Int-s2-PV} with that of using it in
a tadpole integral calculated with a hard momentum cutoff, \eqref{eq:L^2-fermi-tadpole}.
The reason for doing so is that Pauli--Villars is not strong enough
to regulate the tadpole integral, and we do not know of any regulator
both strong enough, and potentially sensitive to independent variations
of $\Lambda_{\mathrm{heavy}}$ and $\Lambda_{\mathrm{light}}$. Nevertheless
the complaint is well-founded: note that one of the finite terms from
Pauli--Villars differs from that given by the hard cutoff --- compare
\eqref{eq:L^2-Int-s2-PV} to \eqref{eq:L^2-fermi-bubble}.

What we believe we \emph{have} shown is that justifiable alterations
to the cutoffs used alter $\delta m^{2}$ by terms of the form $\gamma p_{-}^{1}/\pi\times\mbox{(rational)}$,
which is the same shape as the extra term in \eqref{eq:ww-total-dm2}.
And also that these do \emph{not} produce terms of the form $p_{-}^{2}\log2/\pi$.
That the cancellation is not perfect is perhaps a limitation of our
implementation of this physical cutoff.

\subsection{`Old' and `new' prescriptions\label{sub:Old-and-new-prescriptions}}

We have discussed above very small changes to the cutoffs, and now
turn to a large one. 

First of all, recall how $c=-\log(2)/2\pi$ was found in \cite{McLoughlin:2008he,Abbott:2010yb}.
There, the energy corrections coming from light modes or heavy modes
alone diverge as follows:\[
\delta E\;\propto\;\log\Lambda_{\mathrm{heavy}}-\log2\Lambda_{\mathrm{light}}\:.\]
If we use the same cutoff $\Lambda$ for both of them, they cancel
leaving a factor of $\log2$. This we call the old sum \cite{Bandres:2009kw,Abbott:2010yb}
which is what was used in \cite{McLoughlin:2008ms,Alday:2008ut,Krishnan:2008zs,McLoughlin:2008he}.
If however we use $2\Lambda$ for heavy and $\Lambda$ for light,
then they cancel precisely, leaving $c=0$. This is called the new
sum,%
\footnote{The new sum is very simple from the algebraic curve perspective, and
thus sometimes goes by this name. Likewise the old sum is a cutoff
on worldsheet mode number, which is simple to implement in that formalism.
But both sums can be implemented using any technology. %
} and was originally suggested by \cite{Gromov:2008fy}, see also \cite{Bandres:2009kw,Abbott:2010yb,Astolfi:2011ju}.

What we are doing above is equivalent to the old sum: we treat all
loop momenta $k$ alike. In order to implement the new sum, we should
change the heavy tadpole integral \eqref{eq:log-L-heavy-boson-tadpole}
as follows: \begin{align}
\mathcal{A}_{T2}=\parbox[top][0.6in][c]{1in}{\fmfreuse{tadpole-heavy}}\;\underset{\mathrm{new}}{\longrightarrow}\;\tilde{\mathcal{A}}_{T2} & =\frac{p_{-}^{2}}{4}\frac{i}{2\pi}\int_{0}^{2\Lambda}dK\frac{-K}{K^{2}-1}\nonumber \\
 & =\mathcal{A}_{T2}-\frac{i}{4}p_{-}^{2}\frac{\log2}{2\pi}\,.\label{eq:change-in-A2-for-new-sum}\end{align}
This change precisely cancels the final finite tadpole total, \eqref{eq:ww-tadpole},
leaving $\tilde{\mathcal{A}}_{T}=0$. This can perhaps be viewed as
taking us to $c=0$, as expected for the new sum. To make this change
in dimensional regularisation, we can write the integral \eqref{eq:ww-AT2-tadpole-s}
in terms of $\tilde{k}=k/2$, and then treat $\tilde{k}$ in the same
way as $k$ for the light mode integrals, obtaining the same effect:
\begin{align*}
I_{1}^{0}(1)=\int\frac{d^{2}k}{(2\pi)^{2}}\frac{1}{k^{2}-1}\quad\underset{\mathrm{new}}{\longrightarrow} & \quad I_{1}^{1}(\tfrac{1}{4})=\int\frac{d^{2}\tilde{k}}{(2\pi)^{2}}\frac{1}{\tilde{k}^{2}-\frac{1}{4}}=I_{1}^{0}(1)+\frac{i}{2\pi}\Big\{\frac{\log\tfrac{1}{4}}{2}-\frac{\log1}{2}\Big\}\\
\Rightarrow\quad\mathcal{A}_{T2}=\frac{p_{-}^{2}}{4}I_{1}^{0}(1)\quad\underset{\mathrm{new}}{\longrightarrow} & \quad\mathcal{A}_{T2}-\frac{i}{4}p_{-}^{2}\frac{\log2}{2\pi}\,.\end{align*}

However this tadpole diagram is not the only place in which the heavy
modes play a role: each bubble diagram also has one heavy propagator.
But there is no obvious way to impose a very different cutoff compared
to the light mode in the same bubble. (Attempting to use the Pauli--Villars
procedure above with $\Lambda_{\mathrm{heavy}}=2\Lambda+\ldots$ does
not work: the most divergent term is then $I_{2}^{s=2}\sim\Lambda^{2}\frac{32}{3}\log2$,
which no longer cancels that from the tadpole.) Thus we see no way
to implement something like the new sum for all terms, tadpole and
bubble. 

This argument is in some sense the inverse of that made by \cite{Astolfi:2011ju},
from much the same data. Their cubic interaction $H_{3}$ is essentially
our $\mathcal{L}_{3}$ written in momentum space, and there is of
course a delta function conserving momentum at the vertex. If both
light modes (say $\omega$ and $\bar{\omega}$) carry momentum $\Lambda$,
then the heavy mode ($y$) can carry momentum up to $2\Lambda$. It
is our understanding that this observation is the heart of that paper's
argument in favour of the `new' prescription. While instead we use
the same vertex to draw the bubble diagram, in which it is difficult
to adjust the heavy and light cutoffs independently, leading us to
the `old' prescription.

\subsection{Modes other than the light boson\label{sub:Modes-other-than-w}}

So far we have discussed in section \ref{sec:A-Detour-through-Cutoffs}
only the $\omega$ particle. Let us first comment on how the issues
of section \ref{sub:Old-and-new-prescriptions} translate to other
cases:
\begin{itemize}
\item For the heavy bosons which we treat in the next section, the observation
that the new sum sets the tadpoles to zero carries over trivially:
the calculation of the tadpole contribution is identical. However,
the bubble diagram contains two light modes, thus our objection to
the new sum does not hold for these. 
\item For the light fermion $\psi$, the tadpole \eqref{eq:psi-heavy-integrals}
also contains the $\Lambda^{2}$-divergent integral $I_{1}^{1}(1)$,
for which \[
I_{1}^{1}(1)\underset{\mathrm{new}}{\longrightarrow}4\: I_{1}^{1}(\tfrac{1}{4})=I_{1}^{1}(1)+\frac{i}{2\pi}\frac{\log\tfrac{1}{4}}{2}.\]
This change is identical to that for $I_{1}^{0}(1)$ above. Then it
is easy to see that while we do not create a divergence, the finite
change (from trying to go to the new sum) is \emph{not} so simple
in this case, i.e. it does not cancel the $\log2$ term in \eqref{eq:psi-total-dm2}.
\end{itemize}
The fact that we cannot cancel the $\log2$ like this for $\psi$
is possibly related to another issue. If we attempt to do the whole
calculation with a momentum cutoff $\Lambda$ (generalising section
\ref{sub:Using-hard-cutoffs}) then we do not get a finite result.
The reason for this is not entirely clear to us, but let us observe
here that the cancellations of $1/\epsilon$ terms which make the
result using dimensional regularisation finite are highly nontrivial
--- see \eqref{eq:psi-total-dm2}, and \eqref{eq:BMN-psi-total} below.

This is true also for the heavy bosons of the next section, where
for instance it is clear that the bubble $\mathcal{A}_{y}$ is finite
\eqref{eq:yy-bubble-x-integral}, but the tadpoles $\mathcal{A}_{T}$
are as for the $\omega$ case, and thus $\Lambda^{2}$ divergent \eqref{eq:L^2-fermi-tadpole}.

\section{Correction to the Heavy Boson Propagators\label{sec:Heavy-Prop}}

For a heavy mode we should start with the following dispersion relation:\[
E_{\mathrm{heavy}}=\sqrt{1+16\, h(\lambda)^{2}\sin^{2}\frac{\pchain}{4}}=2E_{\mathrm{light}}\left(\frac{\pchain}{2}\right).\]
This is simply the energy of two superimposed light modes, written
in terms of their \emph{total} momentum. This relation was confirmed
to hold at one loop in the giant magnon regime $\pchain\sim1$ in
\cite{Abbott:2010yb}.%
\footnote{This giant magnon, in an $RP^{2}$ subspace, is a nonlinear superposition
of two elementary magnons, as was shown by \cite{Hollowood:2009sc}.
When discussing giant magnons it is often convenient to write $E=\sqrt{1+8\lambda\sin(p'/2)}$,
where $p'$ is the momentum of one of the constituent magnons, not
the total. (See for instance section 4.3 of \cite{Abbott:2009um}.) %
} Expanding $E^{2}$ in $1/\sqrt{\lambda}$ exactly as for the light
mode case, and in particular using the same normalisation $p_{1}=\sqrt{\lambda/2}\:\pchain$,
we obtain: \begin{equation}
p_{0}^{2}-p_{1}^{2}=1+\left(\frac{c\, p_{-}^{2}}{\sqrt{2\lambda}}-\frac{p_{-}^{4}}{384\lambda}\right)+\mathcal{O}\Big(\frac{1}{\sqrt{\lambda}}\Big).\label{eq:heavy-disp-rel-expanded}\end{equation}
The mass correction $\delta m^{2}$ is again the term in brackets.
Its first term, which we aim to compute here, should be identical
to that for the light modes.

The calculation of course uses a similar geometric series to \eqref{eq:light-geom-series},
although this time $\delta m^{2}=i\gamma\mathcal{A}$ since $y$ and
$z_{i}$ (unlike $\omega^{\alpha}$) are canonically normalised. 

After working out the corrections for fields $y$ and $z_{i}$ we
discuss the issue of their stability in section \ref{sub:Breaking-of-mass-coincidence}.

\subsection{Diagrams correcting $\left\langle yy\right\rangle$}

Recall that for the light boson, the bosonic tadpoles came from \eqref{eq:L4BB-NFS}
and the fermionic ones from the last line of \eqref{eq:L4BF-NFS},
which is this: \begin{align*}
\mathcal{L}_{\mathrm{tad}} & =\frac{-1}{8}\Big[\partial_{-}z_{i}\partial_{-}z_{i}+(\partial_{-}y)^{2}+\partial_{-}\bar{\omega}^{\alpha}\partial_{-}\omega_{\alpha}\Big]\Big(i\bar{\psi}_{-b}\overleftrightarrow{\partial_{+}}\psi_{-}^{b}+\bar{\psi}_{-b}\psi_{+}^{b}+\bar{\psi}_{+b}\psi_{-}^{b}\Big).\end{align*}
Both of these terms share the same first factor, and by looking at
this factor it is clear that we can re-use exactly the same diagrams
when the external particle is $y$. The only change is a factor of
2 because $y$ is real while $\omega_{\alpha}$ was complex, but since
we now have $\delta m^{2}=i\gamma\mathcal{A}$ (without a 2), the
correction $\delta m^{2}$ is unchanged: \begin{align*}
\mathcal{A}_{T'} & =\parbox[top][0.8in][c]{1in}{\fmfreuse{tadpole-yy}}+\parbox[top][0.7in][c]{1in}{\fmfreuse{tadpole-yy-heavy}}=2\mathcal{A}_{T}\\
 & =\frac{i}{2}p_{-}^{2}\frac{\log2}{2\pi}\\
 & \qquad\qquad\qquad\Rightarrow\:\delta m_{T}^{2}=-\frac{1}{2}\gamma p_{-}^{2}\frac{\log2}{2\pi}\quad\mbox{the same as }\eqref{eq:ww-tadpole}.\end{align*}

There is only one bubble diagram to draw, which has two $\omega_{\alpha}$
particles in the loop. The contribution is as follows: \begin{align}
\mathcal{A}_{y} & =\quad\parbox[top][0.8in][c]{1.5in}{\fmfreuse{bubble-yy}},\qquad\mbox{now with }q_{\mu}=k_{\mu}-p_{\mu}\nonumber \\
 & =i^{2}2\int\frac{d^{2}k}{(2\pi)^{2}}\:\frac{2i}{(k^{2}-\frac{1}{4})}\:\frac{2i}{(q^{2}-\frac{1}{4})}\left(\frac{k_{-}+q_{-}}{8}\right)^{2}\displaybreak[0]\nonumber \\
 & =\frac{1}{8}\int_{0}^{1}dx\int\frac{d^{2}\ell}{(2\pi)^{2}}\,\frac{\left(2\ell_{-}+(1-2x)p_{-}\right)^{2}}{\left[\ell^{2}-\Delta'(x)\right]^{2}}\qquad\mbox{where}\quad\begin{aligned}\ell & =k-(1-x)p\quad\mbox{as before,}\\
\Delta' & =x^{2}-x+\tfrac{1}{4}\quad=\tfrac{1}{4}\Delta(4x)\end{aligned}
\displaybreak[0]\nonumber \\
 & =\frac{1}{8}p_{-}^{2}\frac{i}{2\pi}\int_{0}^{1}dx\,\frac{(1-2x)^{2}}{2\Delta'(x)}\label{eq:yy-bubble-x-integral}\\
 & =\frac{i}{8\pi}p_{-}^{2}\:.\nonumber \end{align}

Thus the total mass shift for the $y$ particle is \[
\delta m^{2}=-\frac{1}{2}\gamma p_{-}^{2}\frac{\log2}{2\pi}-\frac{1}{8\pi}\gamma p_{-}^{2}\,.\]
As for the light boson $\omega$ in \eqref{eq:ww-total-dm2}, there
is an extra term not expected from the dispersion relation \eqref{eq:heavy-disp-rel-expanded}.
This now comes from the only bubble diagram possible, rather than
the addition of a bosonic and a fermionic bubble, and thus it is difficult
to imagine a cancellation of this term.

\subsection{Diagrams correcting $\left\langle z_{i}z_{j}\right\rangle $}

The tadpole corrections are completely identical to those for the
$\left\langle yy\right\rangle $ case above, by the same argument
used there.

The only bubble diagram has $\psi$ fields in the loop: \[
\mathcal{A}_{z}=\quad\parbox[top][0.8in][c]{1.4in}{\fmfreuse{bubble-zz}}\]
Since the terms in $\mathcal{L}_{3}$ are written with $Z_{b}^{a}$
instead of $z_{i}$, it is easiest to work in terms of these, and
so we write $\mathcal{A}_{z}=2\mathcal{A}_{01}^{10}$ and calculate
the latter. Here is the integral:\begin{align*}
\mathcal{A}_{01}^{10} & =-i^{2}\int\frac{d^{2}k}{(2\pi)^{2}}\:\frac{B}{(k^{2}-\frac{1}{4})(q^{2}-\frac{1}{4})}\\
 & =\int_{0}^{1}dx\int\frac{d^{2}\ell}{(2\pi)^{2}}\,\frac{B_{00}+\ell_{+}\ell_{-}B_{11}+(\ell_{+}\ell_{-})^{2}B_{22}}{\left[\ell^{2}-\Delta'(x)\right]^{2}},\qquad\Delta'=x^{2}-x+\tfrac{1}{4}\end{align*}
where as before $B=\sum_{s,s'}\ell_{+}^{s}\ell_{-}^{s'}B_{ss'}$,
and we use the same $\Delta'(x)$ as for the $\left\langle yy\right\rangle $
case above. After using $p_{+}p_{-}\approx1$, the coefficients needed
are:\begin{align*}
B_{00} & =-\frac{1}{8}\left(2x-1\right)^{2}\left(3x^{2}-3x-2\right)p_{-}^{2}\\
B_{11} & =\frac{-22x^{2}+22x-3}{4}p_{-}^{2}\\
B_{22} & =-p_{-}^{2}\:.\end{align*}
Expanding using \eqref{eq:dimreg-general-integral}, once again the
logarithmic and quadratic divergences cancel within the bubble term.
The final result is this:\[
\mathcal{A}_{z}=2\left(-\frac{ip_{-}^{2}}{8\pi}\right).\]

The total mass shift for the $z$ particles is thus\[
\delta m^{2}=-\frac{1}{2}\gamma p_{-}^{2}\frac{\log2}{2\pi}+\frac{1}{4\pi}\gamma p_{-}^{2}\,.\]

\subsection{Breaking of mass co-incidence\label{sub:Breaking-of-mass-coincidence}}

The heavy modes have $m=1$, precisely twice the light modes' $m=\frac{1}{2}$.
And the quantum numbers work out that there is a pair of light modes
which carries all the same indices as each heavy mode.%
\footnote{As can be seen from the interaction terms in \eqref{eq:L3-NFS}. The
decomposition of course matches that used in the algebraic curve for
the construction of off-shell frequencies \cite{Gromov:2008bz,Bandres:2009kw,Abbott:2010yb}.%
} One can ask how this gets modified at one loop, and we begin with
some rather simple observations: 
\begin{enumerate}
\item By expanding the dispersion relation we had that, for both heavy and
light modes, \[
\delta m^{2}=c\frac{p_{-}^{2}}{\sqrt{2\lambda}}+\mbox{two loops}.\]
This implies $2\times\Delta m_{\mathrm{light}}=2\delta m^{2}$, and
$\Delta m_{\mathrm{heavy}}=\frac{1}{2}\delta m^{2}$. Using $c=-\log2/2\pi\approx-0.11<0$,
we conclude that after these corrections the heavy mode is \emph{more}
massive than a pair of light modes. Thus it is kinematically allowed
to decay into two light modes.
\item By direct calculation, we find in addition to this some other terms.
These are not large enough to alter the conclusion. Here are the numbers
after including these terms:\begin{align*}
2\:\Delta m_{\omega} & =-0.23\:\gamma p_{-}^{2}\\
2\:\Delta m_{\psi} & =-0.11\:\gamma p_{-}^{2}\end{align*}
and for heavy modes,\begin{align*}
\qquad\Delta m_{y} & =-0.13\:\gamma p_{-}^{2}\\
\Delta m_{z} & =-0.071\:\gamma p_{-}^{2}\:.\:\end{align*}
The decays $y\to\omega+\bar{\omega}$ and $z\to\psi+\bar{\psi}$ are
thus still allowed.
\item If $c=0$, then we expect that $\delta m^{2}=0$, and so at this order
the co-incidence of masses remains unbroken. 
\end{enumerate}
This issue was investigated in a more subtle way by Zarembo in \cite{Zarembo:2009au},
and this was continued in \cite{Sundin:2009zu}. The idea is this:
instead of working exactly at the pole of the original propagator,
$p^{2}=1$ for a heavy mode, we can make an expansion in $p^{2}-1$,
and look at the effect of the second term. 

If we assume that $p^{2}<1$, then the following integral is real:\begin{align*}
\int_{0}^{1}dx\frac{(1-2x)^{2}}{\frac{1}{4}-x(1-x)p^{2}} & =\frac{4}{p^{2}}-\frac{4\sqrt{1-p^{2}}}{p^{3}}\arcsin(p)\\
 & =4-2\pi\sqrt{1-p^{2}}+8(1-p^{2})+\mathcal{O}(1-p^{2})^{3/2}.\end{align*}
This integral is the generalisation away from $p^{2}=1$ of that appearing
in the $\left\langle yy\right\rangle $ bubble \eqref{eq:yy-bubble-x-integral}.
It comes with a factor $i/2\pi$ in $\mathcal{A}$ and a further factor
$i\gamma$ in $\delta m^{2}$ --- in all a minus sign. Thus the term
$4$ here contributes to $\delta m^{2}<0$, ensuring $p^{2}<1$ self-consistently.
In our full calculation there is also the tadpole contribution (proportional
to $\log2$), but this too is negative, as noted above. The conclusion
is that even including the expansion in $p^{2}-1$, the mass corrections
are all real, and so the pole of the corrected propagator $G_{y}(p)$
remains on the real axis.

While this integral diverges for real $p^{2}>1$, the expansion given
does hold for complex $p^{2}$. One might wonder whether, if something
cancelled the leading term 4, the second term might amount to a complex
mass correction. The clearest way (we can think of) to investigate
this as follows. Define some coefficients $B,C,D,F$ like this: \begin{align*}
p^{2} & =1+\delta m^{2},\\
\delta m^{2} & =-\gamma\left[B+C\sqrt{1-p^{2}}+D(1-p^{2})+\mathcal{O}(1-p^{2})^{3/2}\right]+\gamma^{2}F+\mathcal{O}(\gamma^{3}).\end{align*}
Then $B>0$ is the simple case already discussed. If $B=0$, then
$C>0$ looks like it will guarantee you a negative mass shift, thus
keeping $\sqrt{1-p^{2}}$ real, while $C<0$ ... is harder to think
about. But we can solve this equation explicitly for $p^{2}$, and
expanding in $\gamma$, we get\[
p^{2}=1-B\gamma\pm\sqrt{B}\, C\:\gamma^{3/2}+\left(-\frac{C^{2}}{2}-BD+F\right)\gamma^{2}+\mathcal{O}(\gamma^{5/2}).\]
Now it seems clear what happens if $B=0$: regardless of the sign
of $C$ the correction is still a real number, but it enters only
at two loops.

\section{Corrections in the BMN Limit\label{sec:BMN}}

In this section we repeat our calculations above using the BMN Lagrangian
of \cite{Sundin:2009zu}. The physical reason for doing so is that
we want to check that the near-flat-space limit and the diagrammatic
calculation of $\delta m^{2}$ commute. 

When taking the near-flat-space limit of our BMN results, in fact
we recover not only the term from the expansion of the dispersion
relation, but also the extra terms which we found before. This indicates
that these extra terms cannot be artefacts of the simpler limit. In
the BMN calculation, we also find some constant terms which are not
visible in the near-flat-space limit, but which also break supersymmetry. 

The approach and the diagrams needed are identical to those for the
near-flat-space limit, but the number of terms involved is much greater.
Consequently we show much less detail here. But this also means that
the cancellation of divergences is more delicate than it was, and
this provides further checks on our work.

\subsection{The dispersion relation}

We can expand the dispersion relation \eqref{eq:dispersion-relation-light}
in the same way as for the near-flat-space case \eqref{eq:light-disp-rel-expanded}.
The normalisation of the worldsheet momentum is exactly the same $\pchain\sqrt{\lambda/2}=p_{1}$,
as this is a statement only about the gauge we are in. But now we
have $\pchain\sim\lambda^{-1/2}$ from \eqref{eq:three-p-limits},
which changes the result to \begin{equation}
E^{2}=\frac{1}{4}+p_{1}^{2}+\frac{2\sqrt{2}}{\sqrt{\lambda}}c\: p_{1}^{2}+\mathcal{O}\Big(\frac{1}{\lambda}\Big).\label{eq:BMN-disp-rel}\end{equation}
This reads $p_{0}^{2}-p_{1}^{2}=\frac{1}{4}+\delta m^{2}$, and thus
we expect to see \[
\delta m^{2}=\frac{2\sqrt{2}}{\sqrt{\lambda}}c\: p_{1}^{2}\:.\]
This time, the correction term $\delta m^{2}$ is order $1/\sqrt{\lambda}$
while $p_{0}^{2}$ and $p_{1}^{2}$ are order 1. (The two-loop term
we wrote in \eqref{eq:light-disp-rel-expanded} is now order $1/\lambda$.) 

For a heavy mode, $E_{\mathrm{heavy}}(p)=2E(p/2)$, we again get $p_{0}^{2}-p_{1}^{2}=1+\delta m^{2}$
with the same $\delta m^{2}$ at this order.

\subsection{Results for light modes}

For the light bosonic propagator we find: \begin{align}
\qquad\qquad\left\langle \bar{\omega}\:\omega\right\rangle :\qquad\delta m^{2}/\gamma & =\frac{9}{16\pi\epsilon}-\frac{3}{4\pi}p_{1}^{2}+\ldots & \mbox{from bubble diagrams} & \qquad\qquad\nonumber \\
 & \quad\;-\frac{9}{16\pi\epsilon}-\frac{\log2}{\pi}p_{1}^{2}+\ldots & \mbox{from tadpoles}\nonumber \\
 & =-\frac{5}{96\pi}-\frac{3}{4\pi}p_{1}^{2}-\frac{\log2}{\pi}p_{1}^{2} & \mbox{in total}.\label{eq:BMN-w-total}\end{align}
The dots for bubble and tadpole contributions are constant (and non-divergent)
terms. If we now write $p_{1}=\frac{1}{2}(p_{+}-p_{-})$ and perform
the worldsheet boost, then we end up with the result obtained starting
from the near-flat-space Lagrangian. Thus the limit commutes with
the calculation. 

However, for the BMN case, we see that there is one additional non-trivial
$1/\epsilon$ cancellation between the bubbles and the tadpoles which
is invisible in the near-flat-space calculation (where all $1/\epsilon$
canceled within each diagram). This provides a further consistency
check. 

For the light fermionic propagator we have: \begin{align*}
\left\langle \bar{\psi}_{-}\psi_{+}\right\rangle :\qquad\delta m^{2}/\gamma & =\frac{7}{64\pi\epsilon}+\frac{1}{8\pi}\Big(\frac{21}{\epsilon}-\frac{21\gamma_{E}}{2}+13\log2-\frac{21\log\pi}{2}\Big)p_{1}^{2}+\ldots\qquad\mbox{bubbles}\\
 & \quad-\frac{7}{64\pi\epsilon}-\frac{1}{8\pi}\Big(\frac{21}{\epsilon}-\frac{21\gamma_{E}}{2}+21\log2-\frac{21\log\pi}{2}\Big)p_{1}^{2}+\ldots\qquad\mbox{tadpoles}.\end{align*}
The coefficients in this depend on which of the four cases $\left\langle \bar{\psi}_{\pm}\,\psi_{\pm'}\right\rangle $
we consider. But for all cases, the total is the same: \begin{equation}
\left\langle \bar{\psi}\,\psi\right\rangle :\qquad\delta m^{2}/\gamma=-\frac{1}{96\pi}-\frac{\log2}{\pi}p_{1}^{2}\,.\qquad\qquad\qquad\qquad\label{eq:BMN-psi-total}\end{equation}
 Again we see a cancellation of $1/\epsilon$ divergences in both
the constant and the $p_{1}^{2}$ terms. And again the result reduces
to the near-flat-space one, \eqref{eq:psi-total-dm2}.

Notice that in both of these results there are extra constant terms.
These are of course invisible in the near-flat-space limit. But like
the extra terms seen there, they are not the same for the boson and
the fermion, thus breaking supersymmetry.

\subsection{Results for heavy bosons}

Starting out with the $y$ mode, we have \begin{align}
\left\langle yy\right\rangle :\qquad\delta m^{2}/\gamma & =\frac{1}{2\pi\epsilon}-\frac{1}{2\pi}p_{1}^{2}+\ldots\qquad\mbox{bubble}\nonumber \\
 & \quad-\frac{1}{2\pi\epsilon}-\frac{\log2}{\pi}p_{1}^{2}+\ldots\qquad\mbox{tadpoles}\nonumber \\
 & =-\frac{1}{4\pi}-\frac{1}{2\pi}p_{1}^{2}-\frac{\log2}{\pi}p_{1}^{2}.\label{eq:BMN-y-total}\end{align}
The complex piece of $\delta m^{2}$ comes from the $x$ integration
in the Feynman parameterisation.

Finally, for the $z_{i}$ modes (transverse $AdS_{4}$ directions)
we have \begin{align}
\left\langle z_{i}z_{j}\right\rangle :\qquad\delta m^{2}/\gamma & =\frac{1}{4\pi\epsilon}-\frac{p_{1}^{2}}{\pi}+\ldots\qquad\mbox{bubble}\nonumber \\
 & \quad-\frac{1}{4\pi\epsilon}-\frac{\log2}{\pi}p_{1}^{2}+\ldots\qquad\mbox{tadpoles}\nonumber \\
 & =-\frac{2}{3\pi}-\frac{p_{1}^{2}}{\pi}-\frac{\log2}{\pi}p_{1}^{2}.\label{eq:BMN-z-total}\end{align}
As with the light propagators, all corrections reduce to those calculated
using the near-flat-space Lagrangian. Thus, at least for two-point
functions, the truncation of the BMN theory to the near-flat-space
limit is one-loop quantum consistent.

\section{Conclusions\label{sec:Conclusions}}

The results of this paper are as follows:
\begin{enumerate}
\item We have computed one-loop corrections to propagators in both the near-flat-space
and BMN limits, and we find that our calculations `commute with the
limit', meaning that the near-flat results are recovered as limits
of the BMN ones. This is the first such comparison in $AdS_{4}\times CP^{3}$,
and is evidence that the near-flat-space limit is a consistent truncation
at one loop. 
\item Comparing our results to the dispersion relation, we expect $\delta m^{2}$
to be proportional to the $c=-\log2/2\pi$ in the expansion of $h(\lambda)$,
and we correctly find this term. We also (in most cases) find some
extra terms. Our results are always of the form\begin{equation}
\begin{aligned}\qquad\qquad\delta m^{2} & =\;-\frac{\log2}{2\pi}2\gamma p_{1}^{2}+U\frac{\gamma p_{1}^{2}}{\pi}+V\frac{\gamma}{\pi} & \mbox{ BMN limit} & \qquad\qquad\\
 & \to\;-\frac{\log2}{2\pi}\frac{\gamma p_{-}^{2}}{2}+U\frac{\gamma p_{-}^{2}}{4\pi} & \mbox{ near-flat-space},\end{aligned}
\label{eq:dm2-summary-U-V-BMN}\end{equation}
where $U$ and $V$ are rational numbers depending on the mode being
studied. 
\item The term with $\log2$ comes from%
\footnote{This statement is strictly true in cases where the bubble alone is
finite. However the argument that we can remove the $\log2$ term
holds for all the bosons (but fails for the light fermion). %
} the tadpole diagrams, and for these diagrams, we can implement the
`new' summation prescription of Gromov and Mikhailov \cite{Gromov:2008fy}.
Doing so changes the result to remove this $\log2$ term perfectly,
corresponding to $c=0$. This agrees with expectations from \cite{Gromov:2008fy,McLoughlin:2008he,Abbott:2010yb,Astolfi:2011ju}.
However we \emph{cannot} make this change of prescription for the
bubble diagrams, since the same loop momentum applies to both heavy
and light modes.
\end{enumerate}

\subsection[Extra terms $U$ and $V$]{Extra terms $U$ and $V$ in (\ref{eq:dm2-summary-U-V-BMN})}

Perhaps the first thing to note is that these terms break supersymmetry.
Most of our calculation is done using dimensional regularisation,
and so it is natural to suspect that this regulator is the culprit.
(Supersymmetry after all is very different in different dimensions
of spacetime.) It is possible that the terms $U$ and $V$ should
be absorbed into a (finite) renormalisation of the bare masses $m=\tfrac{1}{2},1$,
restoring supersymmetry \cite{Bellucci:1986mz}. In this case the
correct attitude towards our calculation is that we have used the
exact answer for the mass shift (from the dispersion relation) to
work out what bare masses must be used in the propagators for future
calculations, done with the same regulator. 

For the light boson, we were able to repeat our calculation using
instead an explicit momentum cutoff (in section \ref{sub:Using-hard-cutoffs})
and we found exactly the same term $U$. In this case, we gave an
argument that using a more precise cutoff on energy might cancel this
term. The energy, unlike the worldsheet momentum, is a gauge-invariant
physical quantity. This led us in section \ref{sub:High-precision-cutoffs}
to consider cutoffs on the Euclidean momentum of $\left|K\right|<\Lambda^{2}-m^{2}$,
with $m^{2}=\frac{1}{4}$ or $1$ for light or heavy modes. Then because
of a cancellation of $\Lambda^{2}$ divergences, the final result
changes by terms of the form $\gamma p_{-}^{2}/\pi\times(\mbox{rational})$.
It is difficult however to implement an independent cutoff on the
the heavy and light modes in the bubble diagram. We gave one rather
Heath Robinson construction, involving two Pauli--Villars regulators,
which manages to cancel half of the extra term $U$, see \eqref{eq:hyperfine-change-ww}. 

The reason that these regulators are needed at all is essentially
that the leading interaction term $\mathcal{L}_{3}$ contains both
left- and right-moving fields, and derivatives. This does not happen
in the $AdS_{5}\times S^{5}$ case, and as a result the analogous
leading diagrams%
\footnote{By this we mean the two-loop `sunset' diagram correcting the propagator
in \cite{Klose:2007rz}, since the leading interaction there is $\mathcal{L}_{4}$. %
} give only finite integrals. 

We stress that these extra terms are not an artefact of the near-flat-space
limit: the exact terms we find there are also obtained by taking the
limit of our BMN results, as in \eqref{eq:dm2-summary-U-V-BMN}. And
further, the extra term $U$ seen in the near-flat case scales like
$p_{-}^{2}$ in the same way as the $\log2$ term, thus there is no
obvious explanation for why they should not be seen in the giant magnon
case \cite{Abbott:2010yb}, although the $V$ terms would be invisible.

\subsection{Group parameterisation}

Finally, it seems that these extra terms may be altered by the choice
of group parameterisation used. A different parameterisation was defined
in \cite{Zarembo:2009au}, replacing our equation \eqref{eq:group-param-factors}
with \[
G=\Lambda(t,\phi)\, e^{\mathbb{X}}\]
where $\mathbb{X}$ contains both bosons and fermions. The resulting
fields will be related in some complicated nonlinear way, but all
\emph{physical} results must agree. Using this parameterisation, $\mathcal{L}_{3}$
was calculated by the authors of \cite{Kalousios:2009ey}, who were
kind enough to share this term with us. This allows us to repeat the
bubble calculation, and again we find an extra term. But the result
is different: for the bubbles correcting the light boson, equivalent
to \eqref{eq:ww-B-bubble} + \eqref{eq:ww-F-bubble}, we now get \begin{equation}
\delta m_{\mathrm{bubble}}^{2}=-\frac{1}{16\pi}\gamma p_{-}^{2}+\frac{7}{128\pi}\gamma p_{-}^{2}=\frac{-1}{128\pi}\gamma p_{-}^{2}\:.\label{eq:bubble-Zarembo-Chris}\end{equation}
This involves a cancellation of $1/\epsilon$ terms in the fermionic
piece which is as nontrivial as those from our parameterisation.%
\footnote{For the BMN case of this we see terms that reduce to these, plus terms
free of $p_{1}$ including $1/\epsilon$ terms, much like those in
section \ref{sec:BMN}.%
}

The quadratic term $\mathcal{L}_{4}$ has not been calculated, thus
we cannot repeat the tadpole calculations. But we note that in the
integrals used, \eqref{eq:dimreg-tadpole-expansions} has no terms
$\frac{1}{\pi}$, and thus the tadpoles never contribute to terms
of the form $\gamma p_{-}^{2}/\pi\times(\mbox{rational})$. So this
result should survive in the complete answer. 

We do not wish to place too much emphasis on \eqref{eq:bubble-Zarembo-Chris},
which depends on many details of the fermions. We can however make
the simpler observation that the bosonic term in $\mathcal{L}_{3}$
is the same in both parameterisations. This term alone makes the bubble
term for the $\left\langle yy\right\rangle $ correction nonzero,
\eqref{eq:yy-bubble-x-integral}, producing an extra term not expected
from the dispersion relation \eqref{eq:heavy-disp-rel-expanded}.

\subsection{Interpretation of the heavy modes}

The heavy modes are something of a puzzle in this example of the AdS/CFT
correspondence. They are not present in the Bethe ansatz description,
although each one can be made as a simple superposition of two light
modes. But from the point of view of classical strings on $AdS_{4}\times CP^{3}$,
the heavy (bosonic) modes are simply 4 of the 8 transverse spatial
dimensions. 

In \cite{Zarembo:2009au} Zarembo began the investigation of what
loop corrections can teach us about this puzzle. With the corrections
that we find, we can make the simpler observation that the heavy mode
becomes heavier than a pair of light modes, and is thus kinematically
allowed to decay. We discussed this further in section \ref{sub:Breaking-of-mass-coincidence}.

\subsection{Parallels to \cite{Astolfi:2011ju}}

This recent paper has some similarities to ours. First of all note
that the $1/R^{2}$ corrections computed there are analogous to the
$1/\sqrt{\lambda}$ corrections computed here, since $1/\sqrt{\lambda}\sim\alpha'/R^{2}$.
Both papers find a cubic interaction (which was absent in the $AdS_{5}\times S^{5}$
case), and while in \cite{Astolfi:2011ju} this is at order $1/R$,
it only contributes at $1/R^{2}$, much like our bubble diagram needing
two cubic vertices. 

Some of the results in both papers are small-$p$ limits of the giant
magnon case studied in \cite{Abbott:2010yb}. In particular both can
see the $-\log2/2\pi$ term seen there, and by altering the prescription
can remove this (although here subject to the limitations mentioned
above). 

However an important difference is that the present paper is strictly
in infinite volume, $J=\infty$, whereas \cite{Astolfi:2011ju} computes
also some finite-$J$ effects. (For instance we believe that equation
(61) there is a L\"{u}scher F-term.)

\subsection*{Acknowledgements}

For discussions of this work, we thank Dmitri Bykov, Valentina Giangreco
M. Puletti, Tristan McLoughlin, Sameer Murthy, Jeff Murugan, Olof
Ohlsson Sax and especially Konstantin Zarembo. We would also like
to thank Olof and Tristan for comments on the manuscript. 

We are very grateful to the authors of \cite{Kalousios:2009ey}, Chrysostomos
Kalousios, Cristian Vergu and Anastasia Volovich, for sharing their
cubic interaction term with us. 

P.S. would also like to thank Nordita for hospitality while working
on this project. M.C.A. also thanks, in TIFR tradition, the people
of India for their support of basic science. 

\appendix

\section{Fermionic Matrices and Notation\label{sec:Fermionic-Matrices}}

A given matrix $M$ in $\mathfrak{osp}(2,2|6)$ can be represented
as a $10\times10$ super matrix as \[
M=\left(\begin{array}{cc}
X & \theta\\
\eta & Y\end{array}\right)\]
 where the bosonic blocks satisfy \[
X^{t}=-\mathbb{C}_{4}\, X\,\mathbb{C}_{4},\quad X^{\dagger}=-\Gamma_{0}\, X\,\Gamma_{0},\qquad Y^{t}=-Y,\quad Y^{\star}=Y,\]
and $X$ correspond to $\mathfrak{sp}(2,2)$ and $Y$ to $\mathfrak{so}(6)$.
The odd blocks are fermionic, in the sense of having Grassmanian matrix
elements, and satisfy \[
\eta=-\theta^{t}\mathbb{C}_{4},\qquad\theta^{\star}=\mathbb{C}_{4}\,\theta.\]
The matrices $\mathbb{C}_{4}$ and $\Gamma_{0}$ are charge conjugation
and $\Gamma$-matrices of the AdS$_{4}$ space and are given explicitly
in \cite{Sundin:2009zu}. After kappa gauge fixing \cite{Bykov:2009jy},
the fermionic $4\times6$ block $\theta$ is given by%
\footnote{Note that we have performed $\theta\rightarrow\Gamma_{0}\theta$ and
a simple rescaling with $1/\sqrt{2}$ compared to \cite{Sundin:2009zu}.%
} \[
\theta=\frac{1}{\sqrt{2}}\left(\begin{array}{cccccc}
\theta_{1,1} & \theta_{1,2} & \theta_{1,3} & \theta_{1,4} & i\,\theta_{1,4} & i\,\theta_{1,3}\\
\theta_{2,1} & \theta_{2,2} & \theta_{2,3} & \theta_{2,4} & i\,\theta_{2,4} & i\,\theta_{2,3}\\
\theta_{2,1}^{\dagger} & \theta_{2,2}^{\dagger} & \theta_{2,3}^{\dagger} & \theta_{2,4}^{\dagger} & -i\,\theta_{2,4}^{\dagger} & -i\,\theta_{2,3}^{\dagger}\\
-\theta_{1,1}^{\dagger} & -\theta_{1,2}^{\dagger} & -\theta_{1,3}^{\dagger} & -\theta_{1,4}^{\dagger} & i\,\theta_{1,4}^{\dagger} & i\,\theta_{1,3}^{\dagger}\end{array}\right).\]
The eight independent components are given by%
\footnote{We will denote the AdS$_{4}$ SU(2) with indices in the set $\{1,2\}$
and the CP$_{3}$ SU(2) with indices in the set $\{3,4\}$. We also
adopt the somewhat sloppy notation $\bar{\chi}=\chi^{\dagger}$.%
} \begin{align}
 & \theta_{1,1}=\frac{1}{\sqrt{2}}\big(\bar{\psi}_{-2}-\bar{\psi}_{+2}+\psi_{-}^{1}+\psi_{+}^{1}\big),\quad\theta_{1,2}=\frac{i}{\sqrt{2}}\big(\bar{\psi}_{-2}-\bar{\psi}_{+2}-\psi_{-}^{1}-\psi_{+}^{1}\big),\\
\nn & \theta_{2,1}=\frac{1}{\sqrt{2}}\big(-\bar{\psi}_{-1}+\bar{\psi}_{+1}+\psi_{-}^{2}+\psi_{-}^{2}\big),\quad\theta_{2,2}=\frac{i}{\sqrt{2}}\big(-\bar{\psi}_{-1}+\bar{\psi}_{+1}-\psi_{-}^{2}-\psi_{-}^{2}\big),\\
\nn & \theta_{1,3}=\frac{1}{2\sqrt{2}}\big(i(s_{-})_{3}^{1}-i(s_{-})_{4}^{1}+(s_{+})_{3}^{1}-(s_{+})_{4}^{1}\big),\quad\theta_{1,4}=\frac{1}{2\sqrt{2}}\big((s_{-})_{3}^{1}+(s_{-})_{4}^{1}-i(s_{+})_{3}^{1}-i(s_{+})_{4}^{1}\big),\\
\nn & \theta_{2,3}=\frac{1}{2\sqrt{2}}\big(i(s_{-})_{3}^{2}-i(s_{-})_{4}^{2}+(s_{+})_{3}^{2}-(s_{+})_{4}^{2}\big),\quad\theta_{2,4}=\frac{1}{2\sqrt{2}}\big((s_{-})_{3}^{2}+(s_{-})_{4}^{2}-i(s_{+})_{3}^{2}-i(s_{+})_{4}^{2}\big).\end{align}
In the Lagrangians \eqref{eq:action-A} and \eqref{eq:action-A-Pi},
we single out the various graded parts of the current, $A=-G^{-1}\, dG$,
using an automorphism $\Omega$ \[
\Omega(M)=\Upsilon\, M\,\Upsilon^{-1},\qquad\Omega(M^{(k)})=i^{k}\, M\]
where $\Upsilon$ is a constant matrix satisfying $\Upsilon^{2}=-1$.
For \eqref{eq:action-A-Pi}, the auxiliary field $\PI$ is given by
\[
\PI=\frac{i}{2}\PI_{+}\Sigma_{+}+\frac{i}{4}\PI_{-}\Sigma_{-}+\PI_{t}\]
where \[
\PI_{t}=\left(\begin{array}{cc}
\frac{i}{2}\PI_{i}^{(z)}\,\Gamma_{i} & 0\\
0 & \frac{1}{2}\PI^{(y)}\, T_{5}+\PI_{i}^{(\omega)}\,\tau_{i}+\bar{\PI}_{i}^{(\bar{\omega})}\,\bar{\tau}_{i}\end{array}\right).\]
Once again, we point to \cite{Sundin:2009zu} for explicit representation
of the various matrices. However, note that a nice property of $\PI$
is that it projects to $M^{(2)}$ for even elements, \[
Str\,\PI\, A^{(2)}=Str\,\PI\, A.\]

\section{Simplifying the Quartic Lagrangian\label{sec:Simplifying-the-Quartic-L}}

After performing the near-flat-space boost from the quartic BMN Lagrangian,
the pure fermionic quartic piece comes out as \begin{align}
\mathcal{L}_{FF} & =-\frac{1}{16}\Big((s_{-})_{a}^{\beta}(s_{-})_{\beta}^{d}(s_{+})_{\gamma}^{a}\partial_{-}(s_{-})_{d}^{\gamma}-(s_{-})_{a}^{\beta}(s_{-})_{\gamma}^{a}(s_{+})_{\beta}^{d}\partial_{-}(s_{-})_{d}^{\gamma}\Big)-\frac{1}{8}\big(\bar{\psi}_{-}\,\psi_{-}\big)^{2}\nonumber \\
 & +\frac{1}{8}\big(\bar{\psi}_{-}\overleftrightarrow{\partial_{-}}\psi_{-}\big)\big(\bar{\psi}_{-}\overleftrightarrow{\partial_{+}}\psi_{-}\big)+\frac{1}{4}\big(\bar{\psi}_{-}\,\partial_{+}{\psi}_{-}\,\bar{\psi}_{-}\,\partial_{-}\psi_{-}+\partial_{+}{\bar{\psi}_{-}}\,\psi_{-}\,\partial_{-}\bar{\psi}_{-}\,\psi_{-}-\bar{\psi}_{-}\,\psi_{-}\,\partial_{-}\bar{\psi}_{-}\,\partial_{+}\psi_{-}\nonumber \\
 & -\partial_{+}{\bar{\psi}_{-}}\,\partial_{-}\psi_{-}\,\bar{\psi}_{-}\,\psi_{-}\big)+\frac{i}{16}\bar{\psi}_{-}\overleftrightarrow{\partial_{-}}\psi_{-}\big(\bar{\psi}_{+}\,\psi_{-}+\bar{\psi}_{-}\,\psi_{+}\big)-\frac{i}{16}\bar{\psi}_{-}\,\psi_{-}\partial_{-}\big(\bar{\psi}_{+}\,\psi_{-}-\bar{\psi}_{-}\,\psi_{+}\big)\displaybreak[0]\nonumber \\
 & +\frac{i}{2}\bar{\psi}_{-}\,\psi_{-}\big(\bar{\psi}_{+}\,\partial_{-}\psi_{-}-\partial_{-}\bar{\psi}_{-}\,\psi_{+}\big)+\frac{3i}{8}\big(\bar{\psi}_{-}\,\partial_{-}\psi_{-}\,\bar{\psi}_{+}\,\psi_{-}-\partial_{-}\bar{\psi}_{-}\,\psi_{-}\,\bar{\psi}_{-}\,\psi_{+}\big)\nonumber \\
 & +\frac{1}{8}\Big[-\partial_{+}({s}_{-})_{\alpha}^{a}(s_{-})_{c}^{\alpha}\,\bar{\psi}_{-a}\,\partial_{-}\psi_{-}^{c}+\frac{1}{2}\partial_{+}({s}_{-})_{\alpha}^{a}(s_{-})_{c}^{\alpha}\,\partial_{-}\bar{\psi}_{-a}\,\psi_{-}^{c}+\frac{1}{2}\partial_{+}({s}_{-})_{\alpha}^{a}\partial_{-}(s_{-})_{c}^{\alpha}\,\bar{\psi}_{-a}\,\psi_{-}^{c}\nonumber \\
 & +\frac{1}{2}(s_{-})_{\alpha}^{a}\partial_{+}({s}_{-})_{c}^{\alpha}\,\bar{\psi}_{-a}\,\partial_{-}\psi_{-}^{c}-(s_{-})_{\alpha}^{a}\partial_{+}({s}_{-})_{c}^{\alpha}\,\partial_{-}\bar{\psi}_{-a}\,\psi_{-}^{c}+\frac{1}{2}\partial_{-}(s_{-})_{\alpha}^{a}\partial_{+}({s}_{-})_{c}^{\alpha}\,\bar{\psi}_{-a}\,\psi_{-}^{c}\displaybreak[0]\nonumber \\
 & -\frac{1}{4}(s_{-})_{\alpha}^{a}(s_{-})_{c}^{\alpha}\,\bar{\psi}_{-a}\,\psi_{-}^{c}+\frac{1}{2}(s_{-})_{\alpha}^{a}(s_{-})_{c}^{\alpha}\,\partial_{+}{\bar{\psi}}_{-a}\,\partial_{-}\psi_{-}^{c}+\frac{1}{2}(s_{-})_{\alpha}^{a}(s_{-})_{c}^{\alpha}\,\partial_{-}\bar{\psi}_{-a}\,\partial_{+}{\psi}_{-}^{c}\nonumber \\
 & -\frac{1}{2}(s_{-})_{\alpha}^{a}\partial_{-}(s_{-})_{c}^{\alpha}\,\bar{\psi}_{-a}\,\partial_{+}{\psi}_{-}^{c}-\frac{1}{2}\partial_{-}(s_{-})_{\alpha}^{a}(s_{-})_{c}^{\alpha}\,\partial_{+}{\bar{\psi}}_{-a}\,\psi_{-}^{c}-2(s_{-})_{\alpha}^{a}(s_{+})_{c}^{\alpha}\,\partial_{-}\bar{\psi}_{-a}\,\psi_{-}^{c}\nonumber \\
 & -2(s_{+})_{\alpha}^{a}(s_{-})_{c}^{\alpha}\,\bar{\psi}_{-a}\,\partial_{-}\psi_{-}^{c}+i\partial_{-}(s_{-})_{\alpha}^{a}(s_{-})_{c}^{\alpha}\,\bar{\psi}_{+a}\,\psi_{-}^{c}-i(s_{-})_{\alpha}^{a}\partial_{-}(s_{-})_{c}^{\alpha}\,\bar{\psi}_{-a}\,\psi_{+}^{c}\Big]\,.\label{eq:L4FF-NFS-initial}\end{align}
The first thing to notice is that, using total derivatives, the piece
quartic in $\psi$ and $s$ and involving one left-mover can be written
as\begin{align}
 & -\frac{1}{24}(s_{-})_{a}^{\beta}(s_{-})_{\beta}^{d}(s_{-})_{d}^{\gamma}\partial_{-}(s_{+})_{\gamma}^{a}+\frac{i}{8}\big(3\,\bar{\psi}_{-}\,\partial_{-}\psi_{-}+\partial_{-}\bar{\psi}_{-}\,\psi_{-}\big)\bar{\psi}_{-}\,\psi_{+}\nonumber \\
 & -\frac{i}{8}\big(3\,\partial_{-}\bar{\psi}_{-}\,\psi_{-}+\bar{\psi}_{-}\,\partial_{-}\psi_{-}\big)\bar{\psi}_{+}\,\psi_{-}-\frac{i}{2}\bar{\psi}_{-}\,\psi_{-}\big(\partial_{-}\bar{\psi}_{+}\,\psi_{-}-\bar{\psi}_{-}\,\partial_{-}\psi_{+}\big)\,.\label{eq:shift1}\end{align}
If we now perform the following redefinitions of the fermions (and
corresponding ones for conjugated $\psi_{\pm}$), \begin{align*}
 & \delta\psi_{+}^{a}=-\frac{1}{2}\psi_{-}^{a}\bar{\psi}_{-}\,\psi_{-},\qquad\delta\psi_{-}^{a}=-\frac{i}{4}\psi_{-}^{a}\big(3\,\partial_{-}\bar{\psi}_{-}\,\psi_{-}+\bar{\psi}_{-}\partial_{-}\psi_{-}\big),\\
 & \delta(s_{+})_{a}^{\gamma}=-\frac{i}{24}(s_{-})_{a}^{\beta}(s_{-})_{\beta}^{d}(s_{-})_{d}^{\gamma},\end{align*}
the mass term in the quadratic Lagrangian (\ref{eq:L2-lightcone})
removes (\ref{eq:shift1}). However, the kinetic term induces additional
terms giving \begin{align*}
 & \mathcal{L}_{FF}=\frac{3}{8}\big(\bar{\psi}_{-}\,\psi_{-}\big)^{2}-\frac{1}{4}\bar{\psi}_{-}\,\psi_{-}\big(\partial_{+}{\bar{\psi}}_{-}\,\partial_{-}\psi_{-}+\partial_{-}\bar{\psi}_{-}\,\partial_{+}{\psi}_{-}\big)-\frac{3}{8}\big(\bar{\psi}_{-}\,\partial_{+}{\psi}_{-}\,\bar{\psi}_{-}\,\partial_{-}\psi_{-}+\partial_{+}{\bar{\psi}}_{-}\,\psi_{-}\,\partial_{-}\bar{\psi}_{-}\,\psi_{-}\\
 & +\partial_{-}\bar{\psi}_{-}\,\psi_{-}\,\bar{\psi}_{-}\,\partial_{+}{\psi}_{-}+\partial_{+}{\bar{\psi}}_{-}\,\psi_{-}\,\bar{\psi}_{-}\,\partial_{-}\psi_{-}\big)-\frac{1}{24}(s_{-})_{a}^{\beta}(s_{-})_{\beta}^{d}(s_{-})_{d}^{\gamma}(s_{-})_{\gamma}^{a}+...\end{align*}
where the dots denotes the unchanged terms mixing $s_{\pm}$ and $\psi_{\pm}$.
The mixing terms linear in left movers \begin{equation}
\frac{1}{4}\Big((s_{+})_{c}^{\alpha}(s_{-})_{\alpha}^{a}\,\partial_{-}\bar{\psi}_{-a}\psi_{-}^{c}-(s_{+})_{\alpha}^{a}(s_{-})_{c}^{\alpha}\,\bar{\psi}_{-a}\,\partial_{-}\psi_{-}^{c}+\frac{i}{2}\bar{\psi}_{+a}\partial_{-}(s_{-})_{\alpha}^{a}(s_{-})_{c}^{\alpha}\,\psi_{-}^{c}-\frac{i}{2}(s_{-})_{\alpha}^{a}\partial_{-}(s_{-})_{c}^{\alpha}\,\bar{\psi}_{-a}\psi_{+}^{c}\Big)\label{eq:shift2}\end{equation}
can be removed by performing a similar field redefinition. That is,
we introduce \begin{align*}
\delta & (s_{-})_{\alpha}^{c}=-\frac{i}{4}(s_{-})_{\alpha}^{a}\big(\partial_{-}\bar{\psi}_{-a}\psi_{-}^{c}+\bar{\psi}_{-}^{c}\partial_{-}\psi_{-a}\big),\qquad\delta\psi_{-}^{a}=\frac{i}{4}\partial_{-}(s_{-})_{\alpha}^{a}(s_{-})_{c}^{\alpha}\,\psi_{-}^{c}\end{align*}
which removes \eqref{eq:shift2} at the cost of additional quartic
terms. Adding it all together gives the $\mathcal{L}_{FF}$ presented
in \eqref{eq:L4FF-NFS-final}.

\section{Expansions in Dimensional Regularisation\label{sec:A-DimReg-Options}}

In this appendix we consider some changes to our calculation which
would produce terms of the form $\gamma p_{-}^{2}/\pi\times(\mbox{rational})$,
thus altering the terms $U$ in \eqref{eq:dm2-summary-U-V-BMN}. While
we believe that what we have done is correct, we include these other
options as a matter of curiosity. 

The integrals we write in \eqref{eq:dimreg-general-integral} for
$s=1,2$ are closely related to these two:%
\footnote{Here we note that \cite{Peskin:1995ev} has a misprint in (A.47),
which is the $s=2$ case of our \eqref{eq:dimreg-general-integral},
as well as in (A.51).%
} \begin{equation}
\begin{aligned}\int\frac{d^{d}k}{(2\pi)^{d}}\:\frac{k_{\mu}k_{\nu}}{\left[k^{2}-\Delta\right]^{n}} & =\frac{-i}{(4\pi)^{d/2}}\:\frac{g_{\mu\nu}}{2}\:\left(\frac{1}{\Delta}\right)^{n-1-d/2}\frac{\Gamma(n-1-\frac{d}{2})}{\Gamma(n)}\\
\int\frac{d^{d}k}{(2\pi)^{d}}\:\frac{k_{\mu}k_{\nu}k_{\xi}k_{\varpi}}{\left[k^{2}-\Delta\right]^{n}} & =\frac{i}{(4\pi)^{d/2}}\:\frac{g_{\mu\nu}g_{\xi\varpi}+g_{\mu\xi}g_{\nu\varpi}+g_{\mu\varpi}g_{\nu\xi}}{4}\:\left(\frac{1}{\Delta}\right)^{1-d/2}\frac{\Gamma(n-2-\frac{d}{2})}{\Gamma(n)}\:.\end{aligned}
\label{eq:untraced-integrals}\end{equation}
Tracing produces the factors of $d/2$ and $\frac{d}{4}(d+2)$, which
in \eqref{eq:dimreg-general-integral} are written as a quotient of
gamma functions:\begin{equation}
\begin{aligned}g_{\mu\nu}\frac{g_{\mu\nu}}{2} & \:=\frac{d}{2}\quad=\frac{\Gamma(\tfrac{d}{2}+1)}{\Gamma(\tfrac{d}{2})}\\
g_{\mu\nu}g_{\xi\varpi}\frac{g_{\mu\nu}g_{\xi\varpi}+g_{\mu\xi}g_{\nu\varpi}+g_{\mu\varpi}g_{\nu\xi}}{4} & \:=\frac{1}{4}d(d+2)\quad=\frac{\Gamma(\tfrac{d}{2}+2)}{\Gamma(\tfrac{d}{2})}\:.\end{aligned}
\label{eq:gamma-d-d(d+2)-factors}\end{equation}
These factors are multiplied by the divergent gamma function $\Gamma(\frac{\epsilon}{2})$
or $\Gamma(-1+\frac{\epsilon}{2})$, and to get the finite terms shown
in \eqref{eq:dimreg-general-integral} we have expanded these (and
the other non-divergent factors) up to order $\epsilon$. This seems
like the most natural and correct thing to do, when starting from
integrals with $k^{2}$ or $(k^{2})^{2}$ in the numerator. There
are however some other options:
\begin{enumerate}
\item [2.]\setcounter{enumi}{2}  We could expand neither of the factors
in \eqref{eq:gamma-d-d(d+2)-factors}. In other words, use both of
the integrals \eqref{eq:untraced-integrals} rather than \eqref{eq:dimreg-general-integral},
and trace them only at the end. This produces changes to the various
$1/\pi$ terms, which (in the BMN case) lead to the following: \[
\delta m^{2}=-\frac{\log2}{\pi}\gamma p_{1}^{2}+\gamma\begin{cases}
-\frac{13}{12\pi}p_{1}^{2}+\frac{19}{64\pi}-\frac{9a}{32\pi}\qquad\mbox{for} & \omega\\
\frac{17}{96\pi}p_{1}^{1}-\frac{35}{768\pi}-\frac{9a}{32\pi} & \psi\\
-\frac{1}{4\pi}p_{1}^{2}-\frac{17}{48\pi}-\frac{i}{4} & y\\
-\frac{11}{12\pi}p_{1}^{2}+\frac{1}{12\pi}-\frac{3a}{4\pi} & z\end{cases}\]
None of these are preferable to our results above. Note the appearance
of the gauge parameter $a$ in some of these, although only in the
constant the term (and not visible in the near-flat limit). 
\item [3.]\setcounter{enumi}{3} We could expand the $(d+2)$ but not the
$d/2$ in \eqref{eq:gamma-d-d(d+2)-factors}. This would mean in some
sense that we are keeping the same indices $\mu\nu$ outside the integral
in all cases --- we use the first integral in \eqref{eq:untraced-integrals}
but trace the second integral once, like this: \begin{align*}
g_{\mu\nu}g_{\xi\varpi}\frac{g_{\mu\nu}g_{\xi\varpi}+g_{\mu\xi}g_{\nu\varpi}+g_{\mu\varpi}g_{\nu\xi}}{4} & =\frac{g_{\mu\nu}(d+2)}{4}\\
\Rightarrow\quad\int\frac{d^{d}k}{(2\pi)^{d}}\:\frac{k_{\mu}k_{\nu}\: k^{2}}{\left[k^{2}-\Delta\right]^{n}}=\frac{i}{(4\pi)^{d/2}} & \:\frac{g_{\mu\nu}(d+2)}{4}\:\left(\frac{1}{\Delta}\right)^{1-d/2}\frac{\Gamma(n-2-\frac{d}{2})}{\Gamma(n)}\:.\end{align*}
This third option leads to the following results: \[
\delta m^{2}=-\frac{\log2}{\pi}\gamma p_{1}^{2}+\gamma\begin{cases}
-\frac{1}{2\pi}p_{1}^{2}+\frac{5}{64\pi}-\frac{9a}{32\pi}\qquad\mbox{for} & \omega\\
\frac{3}{8\pi}p_{1}^{1}-\frac{1}{384\pi}-\frac{9a}{32\pi} & \psi\\
-\frac{1}{4\pi}p_{1}^{2}-\frac{17}{48\pi}-\frac{i}{4} & y\\
-\frac{3}{4\pi}p_{1}^{2}-\frac{5}{24\pi}-\frac{3a}{4\pi} & z\end{cases}\]
Again these are not any better than our initial results. 
\end{enumerate}
\begin{small}\bibliographystyle{my-JHEP-3}
\addcontentsline{toc}{section}{\refname}\bibliography{/Users/me/Documents/Papers/complete-library-processed,complete-library-processed}
\end{small}

\end{document}